\def\isarxiv{1} %%% for ML conference submission version, we comment this line

\ifdefined\isarxiv
\documentclass[11pt]{article}

\usepackage[numbers]{natbib}

\else
\documentclass{article}
\usepackage{neurips_2022}
\fi

\usepackage{amsmath}
\usepackage{amsthm}
\usepackage{amssymb}
\usepackage{algorithm}
\usepackage{subfig}
\usepackage{algpseudocode}
\usepackage{graphicx}
\usepackage{grffile}
\usepackage{wrapfig,epsfig}
\usepackage{url}
\usepackage{xcolor}
\usepackage{epstopdf}
\usepackage[hypertexnames=false]{hyperref}

\usepackage{bbm}
\usepackage{dsfont}

\allowdisplaybreaks

\ifdefined\isarxiv

\usepackage{tikz}
\usepackage{hyperref}  %%% arxiv don't allow this.
\hypersetup{colorlinks=true,citecolor=blue,linkcolor=blue} %%% Zhao : maybe we should comment this in submission.
\usetikzlibrary{arrows}
\usepackage[margin=1in]{geometry}

\else

\usepackage{microtype}
\usepackage{hyperref}
\definecolor{mydarkblue}{rgb}{0,0.08,0.45}
\hypersetup{colorlinks=true, citecolor=mydarkblue,linkcolor=mydarkblue}

\fi
\graphicspath{{./figs/}}
\usepackage{cleveref}

\newtheorem{theorem}{Theorem}[section]
\newtheorem{lemma}[theorem]{Lemma}
\newtheorem{definition}[theorem]{Definition}

\newtheorem{conjecture}[theorem]{Conjecture}

\newtheorem{fact}[theorem]{Fact}
\newtheorem{remark}[theorem]{Remark}
\newtheorem{claim}[theorem]{Claim}

\newcommand{\wt}{\widetilde}
\newcommand{\ov}{\overline}

\newcommand{\R}{\mathbb{R}}

\newcommand{\Tmat}{{\cal T}_{\mathrm{mat}}}

\DeclareMathOperator{\nnz}{nnz}

\DeclareMathOperator{\diag}{diag}

\DeclareMathOperator{\Att}{\mathsf{Att}}
\DeclareMathOperator{\ct}{\mathrm{ct}}

\makeatletter
\newcommand*{\RN}[1]{\expandafter\@slowromancap\romannumeral #1@}
\makeatother

\usepackage{lineno}

\begin{document}

\ifdefined\isarxiv

\date{}

\title{Algorithm and Hardness for Dynamic Attention Maintenance in Large Language Models}
\author{
Jan van den Brand\thanks{\texttt{vdbrand@gatech.edu}. Georgia Institute of Technology.}
\and
Zhao Song\thanks{\texttt{zsong@adobe.com}. Adobe Research.}
\and
Tianyi Zhou\thanks{\texttt{t8zhou@ucsd.edu}. University of California San Diego.}
}

\else

\title{Intern Project} 
\maketitle 
\fi

\ifdefined\isarxiv
\begin{titlepage}
  \maketitle
  \begin{abstract}

Large language models (LLMs) have made fundamental changes in human life. The attention scheme is one of the key components over all the LLMs, such as BERT, GPT-1, Transformers, GPT-2, 3, 3.5 and 4. Inspired by previous theoretical study of static version of the attention multiplication problem [Zandieh,  Han, Daliri, and Karbasi arXiv 2023, Alman and Song arXiv 2023]. In this work, we formally define a dynamic version of attention matrix multiplication problem. 
There are matrices $Q,K, V \in \mathbb{R}^{n \times d}$, they represent query, key and value in LLMs. In each iteration we update one entry in $K$ or $V$. In the query stage, we receive $(i,j) \in [n] \times [d]$ as input, and want to answer $(D^{-1} A V)_{i,j}$, where $A:=\exp(QK^\top) \in \mathbb{R}^{n \times n}$ is a square matrix and $D := \mathrm{diag}(A {\bf 1}_n) \in \mathbb{R}^{n \times n}$ is a diagonal matrix. Here ${\bf 1}_n$ denote a length-$n$ vector that all the entries are ones.

We provide two results: an algorithm and a conditional lower bound.
\begin{itemize}
    \item On one hand, inspired by the lazy update idea from [Demetrescu and Italiano FOCS 2000, Sankowski FOCS 2004, Cohen, Lee and Song STOC 2019, Brand SODA 2020], 
    we provide a data-structure that uses $O(n^{\omega(1,1,\tau)-\tau})$ amortized update time, 
    and $O(n^{1+\tau})$ worst-case query time.
    \item On the other hand, show that unless the hinted matrix vector multiplication conjecture [Brand, Nanongkai and Saranurak FOCS 2019] is false, there is no algorithm that can use both $O(n^{\omega(1,1,\tau) - \tau- \Omega(1)})$ amortized update time, and $O(n^{1+\tau-\Omega(1)})$ worst query time.  
\end{itemize}
In conclusion, our algorithmic result is conditionally optimal unless hinted matrix vector multiplication conjecture is false.

One notable difference between prior work [Alman and Song arXiv 2023] and our work is, their techniques are from the area of fine-grained complexity, and our techniques are not. Our algorithmic techniques are from recent work in convex optimization, e.g. solving linear programming. Our hardness techniques are from the area of dynamic algorithms.

  \end{abstract}
  \thispagestyle{empty}
\end{titlepage}

{\hypersetup{linkcolor=black}
%\tableofcontents
}
\newpage

\else

\begin{abstract}

\end{abstract}

\fi

\section{Introduction}

Large language models (LLMs) such as Transformer \cite{vsp+17}, BERT \cite{dclt18}, GPT-3 \cite{bmr+20}, PaLM \cite{cnd+22}, and OPT \cite{zrg+22} offer better results when processing natural language compared to smaller models or traditional techniques. These models possess the capability to understand and produce complex language, which is beneficial for a wide range of applications like language translation, sentiment analysis, and question answering. LLMs can be adjusted to multiple purposes without requiring them to be built from scratch. A prime example of this is ChatGPT, a chat software developed by OpenAI utilizing GPT-3's potential to its fullest.
 GPT-4 \cite{openai23}, the latest iteration, has the potential to surpass the already impressive abilities of GPT-3, including tasks such as language translation, question answering, and text generation. As such, the impact of GPT-4 on NLP could be significant, with new applications potentially arising in areas like virtual assistants, chatbots, and automated content creation.

The primary technical foundation behind LLMs is the attention matrix \cite{vsp+17,rns+18,dclt18,bmr+20}. Essentially, an attention matrix is a square matrix with corresponding rows and columns representing individual words or ``tokens," and entries indicating their correlations within a given text. This matrix is then utilized to gauge the essentiality of each token in a sequence, relative to the desired output. As part of the attention mechanism, each input token is assigned a score or weight based on its significance or relevance to the current output, which is determined by comparing the current output state and input states through a similarity function.

More formally, the attention matrix can be expressed as follows: Suppose we have two matrices, $Q$ and $K$, comprising query and key tokens respectively, where $Q \in \R^{n \times d}$ and $K \in \R^{n \times d}$. 
The attention matrix is a square $n \times n$ matrix denoted by $A$ that relates the input tokens in the sequence.
After normalizing using the softmax function, each entry in this matrix quantifies the attention weight or score between a specific input token (query token $Q$) and an output token (key token $K$). Notably, entries along the diagonal reflect self-attention scores, indicating the significance of each token in relation to itself.

When modeling long sequences with large $n$, the most significant hindrance to accelerating LLM operations is the duration required for carrying out attention matrix calculations \cite{kkll20,wlk+20}. These calculations involve multiplying the attention matrix $A$ with another value token matrix $V \in \R^{n \times d}$. In \cite{wlk+20}, they demonstrate that the self-attention mechanism can be approximated by a low-rank matrix. They propose a new self-attention mechanism and used it in their Linformer model. In \cite{kkll20}, they replace dot-product attention with one that uses locality-sensitive hashing, which also improves the time complexity.

Furthermore, the static attention computation and approximation has been studied by \cite{as23} from both algorithmic and hardness perspectives.
However, in practice, the attention matrix needs to be trained and keeps changing. 
In this work, we study the dynamic version of the attention computation problem.
By using a dynamic approach, the attention weights can be updated on-the-fly as new information is introduced, enabling the model to adapt more effectively to changes in the input. 
This is particularly beneficial in cases where the input data is highly dynamic and subject to frequent changes, such as in natural language processing applications where the meaning and context of words and phrases can be influenced by the surrounding text.

\begin{figure}[!ht]
    \centering
    \includegraphics[width = 0.48\textwidth]
    {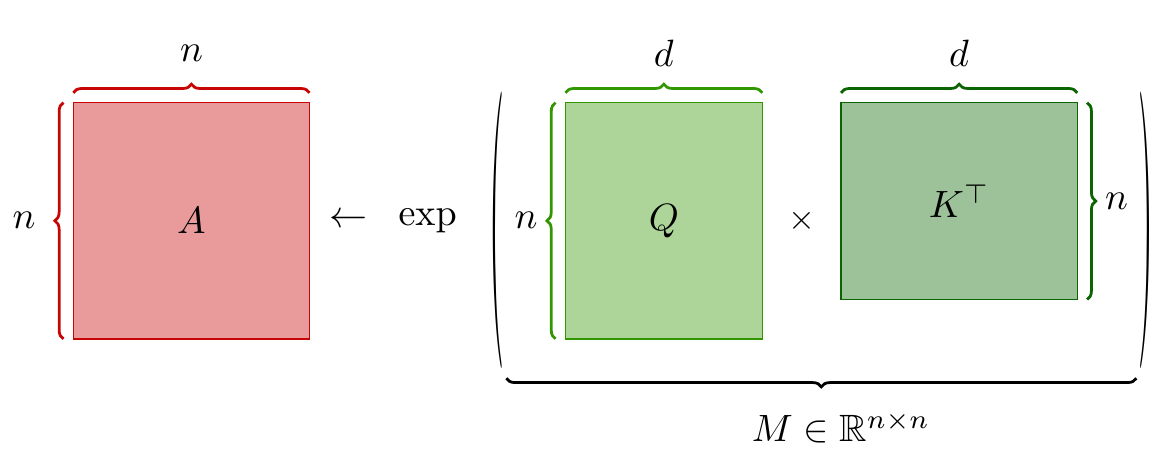}
    \hspace{1mm}
    \includegraphics[width = 0.48\textwidth]{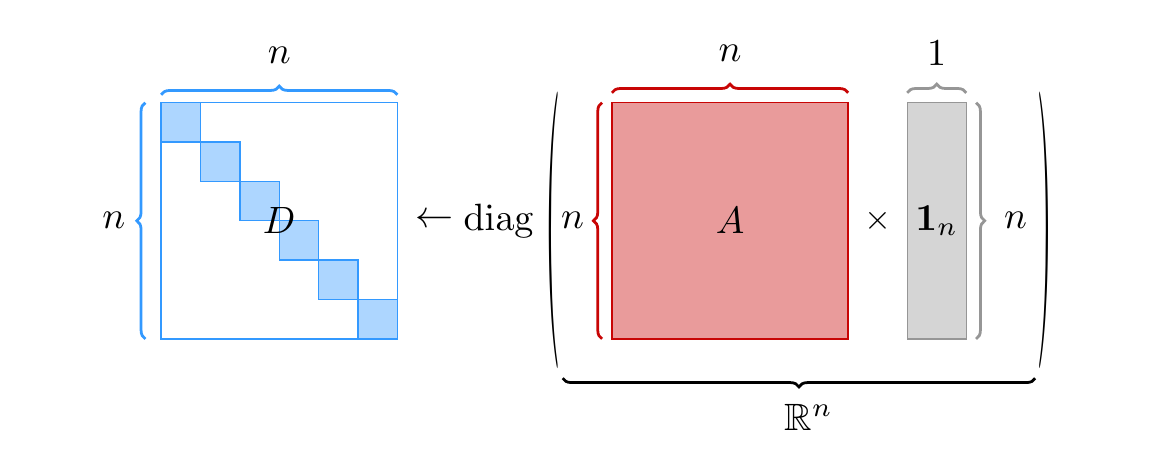}
    \caption{Computation of the attention matrix $A= \exp(Q K^\top)$ and the diagonal matrix $D \in \R^{n \times n}$ (defined in Definition \ref{def:att_mul}). Here $\exp()$ is the entry-wise function.}
    \label{fig:def_A}
\end{figure}

\begin{figure}[!ht]
    \centering
    \includegraphics[width = 0.55\textwidth]{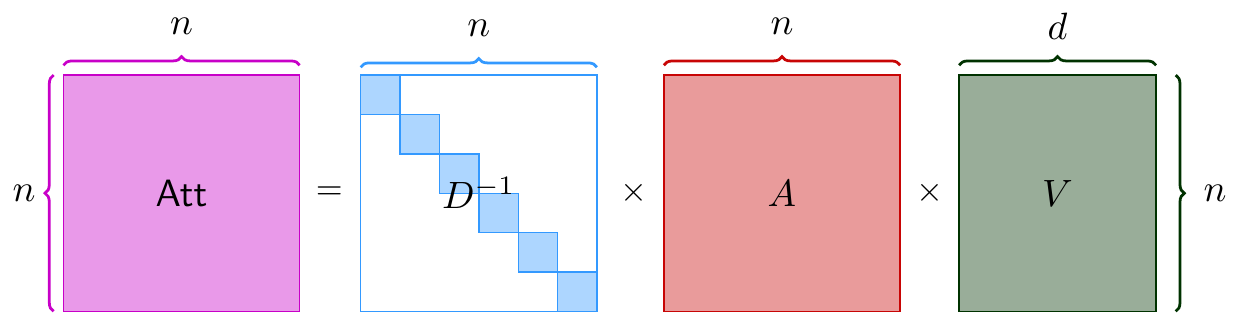}
    \caption{Computation of the target matrix $\Att(Q,K,V) = D^{-1} A V$ (defined in Definition \ref{def:att_mul})}
    \label{fig:def_Att}
\end{figure}

Following the prior work \cite{zhdk23,as23}, we formally define the standard attention computation problem as follows. To distinguish their standard model with the dynamic version studied in this paper, we call the problem defined in \cite{zhdk23,as23} ``static'' version of attention multiplication. 
Another major difference between previous work \cite{zhdk23,as23} and our work is that they studied an approximate version, whereas we study the exact version.
\begin{definition}[Static Attention Multiplication]\label{def:att_mul}
Given three matrices $Q, K, V\in \R^{n \times d}$, 
we define attention computation
\begin{align*}
\Att(Q,K,V) = D^{-1} A V
\end{align*}
where square matrix $A \in \R^{n \times n}$ and diagonal matrix $D \in \R^{n \times n}$ are 
\begin{align*}
A:= \exp(Q K^\top) , ~~~ D := \diag( A {\bf 1}_n )
\end{align*}
Here we apply the $\exp(\cdot)$ function entry-wise\footnote{For a matrix $M \in \R^{n \times n}$, following the transformer literature, we use $\exp(M)_{i,j} := \exp(M_{i,j})$. Our $\exp(\cdot)$ is not the matrix exponential from matrix Chernoff bound literature \cite{glss18}.}. We use ${\bf 1}_n$ to denote a length-$n$ vector where all the entries are ones. The $\diag()$ function is taking a length-$n$ vector as input and outputs an $n \times n$ diagonal matrix by copying that vector on the diagonal of the output matrix. See Figure~\ref{fig:def_A} and Figure \ref{fig:def_Att} for an illustration.
\end{definition}


In applied LLMs training, the model parameters are changing slowly during training \cite{clp+21}. Thus, it is worth considering the dynamic version of Attention multiplication problem. Next, we formally define the ``dynamic'' or ``online'' version of attention multiplication problem, we call it $\mathsf{ODAMV}$\footnote{The name of our problem is inspired by a well-known problem in theoretical computer science which is called {\bf O}nline {\bf M}atrix {\bf V}ector multiplication problem ($\mathsf{OMV}$)  \cite{hkns15,lw17, ckl18}.}. For consistency of the discussion, we will use the word ``online'' in the rest of the paper.

\begin{definition}[$\mathsf{ODAMV}(n,d)$]\label{def:AHDAMV_informal}
The goal of \textbf{\textit{O}}nline \textbf{\textit{D}}iagonal-based normalized \textbf{\textit{A}}ttention \textbf{\textit{M}}atrix \textbf{\textit{V}}ector multiplication problem $\mathsf{ODAMV}(n,d)$ is to design a data-structure that satisfies the following operations:
\begin{enumerate}
    \item \textsc{Init}: Initialize on three $n \times d$ matrices $Q$, $K$, $V$.
    \item \textsc{Update}: Change any entry of $K$, or $V$.
    \item \textsc{Query}: For any given $i \in [n]$, $j\in[d]$, return $(D^{-1}\exp(QK^\top)V)_{i,j}$.
    \begin{itemize}
        \item Here $D := \diag( \exp(QK^\top) {\bf 1}_n ) \in \R^{n \times n}$ is a positive diagonal matrix.
        \item Here $[n]$ denotes the set $\{1,2, \cdots, n\}$.
    \end{itemize}
\end{enumerate}
\end{definition}

In this paper, we first propose a data-structure that efficiently solves the $\mathsf{ODAMV}$ problem (Definition \ref{def:AHDAMV_informal}) by using lazy update techniques. 
When then complement our result by a conditional lower bound. 
On the positive side, we use lazy update technique in the area of dynamic algorithms to provide an upper bound. 
In the area of theoretical computer science, it is very common to assume some 
conjecture 
in complexity when proving a lower bound. For example, $\mathsf{P} \neq \mathsf{NP}$, (strong) exponential time hypothesis, orthogonal vector and so on. 
To prove our conditional lower bound, 
we use a 
conjecture which is called {\bf H}inted {\bf M}atrix {\bf V}ector multiplication ($\mathsf{HMV}$) conjecture \cite{bns19}.
On the negative side, we show a lower bound of computing solving $\mathsf{ODAMV}$ assuming the $\mathsf{HMV}$ conjecture holds.

\subsection{Our Results}

We first show our upper bound result making use of the lazy update strategy.
\begin{theorem}[Upper bound, informal version of Theorem \ref{thm:fast}]\label{thm:fast_informal}
For any constant $a \in (0,1]$. Let $d = O(n)$.
There is a dynamic data structure that uses $O(n^2 )$ space and supports the following operations:
\begin{itemize}
    \item \textsc{Init}$(Q,K,V) $. It runs in $O( \Tmat(n,n,n))$ time.\footnote{We use $\Tmat(n,d,m)$ to denote the time of multiplying a $n \times d$ matrix with another $d \times m$ matrix. 
    } 
    \item \textsc{UpdateK}$(i \in [n], j \in [d], \delta \in \R )$. This operation updates one entry in $Q$, and it runs in $O(\Tmat(n,n^a, n) / n^a )$ amortized time. 
    \item \textsc{UpdateV}$(i \in [n], j \in [d], \delta \in \R)$. This operation takes same amortized time as $Q$ update.
    \item \textsc{Query}$(i \in [n],j \in [d])$. This operation outputs $(D^{-1} (\exp(QK^\top)) V)_{i,j}$ and takes $O(n^a)$ worst-case time.
\end{itemize}
\end{theorem}

Our second result makes use of a variation of the popular online matrix vector multiplication ($\mathsf{OMV}$) conjecture 
which is called hinted matrix vector multiplication conjecture (see Definition \ref{def:hintedMv} and \cite{bns19}). Next, we  present a lower bound for the problem of dynamically maintaining the attention computation $\Att(Q,K,V)$.

\begin{lemma}[Lower bound, informal version of Lemma \ref{lem:lowerbound_D}]
Assuming $\mathsf{HMV}$ conjecture is true. 
For every constant $0 < \tau \le 1$, there is no algorithm that solve $\mathsf{ODAMV}(n,d)$ problem (see formal version in Definition~\ref{def:AHDAMV}) with 
\begin{itemize}
    \item polynomial initialization time, and
    \item amortized update time $O(\Tmat(n, n^\tau, d) / n^{\tau + \Omega(1)})$, and
    \item worst query time $O(n^{\tau - \Omega(1)} )$.
\end{itemize}
\end{lemma}

\subsection{Related Work}

\paragraph{Static Attention Computation}
A recent work by Zandieh, Han, Daliri, and Karbasi~\cite{zhdk23} was the first to give an algorithm with provable guarantees for approximating the attention computation. Their algorithm makes use of locality sensitive hashing (LSH) techniques \cite{ckns20}.
They show that the computation of partition functions in the denominator of softmax function can be reduced to a variant of the kernel density estimation (KDE) problem, and an efficient KDE solver can be employed through subsampling-based swift matrix products. They propose the KDEformer which can approximate the attention within sub-quadratic time and substantiated with provable spectral norm bounds. In contrast, earlier findings only procure entry-wise error bounds. Based on empirical evidence, it was confirmed that KDEformer outperforms other attention approximations in different pre-trained models, in accuracy, memory, and runtime.

In another recent work \cite{as23}, they focus on the long-sequence setting with $d= O(\log n)$. The authors established that the existence of a fast algorithm for approximating the attention computation is dependent on the value of $B$, given the guarantees of $\|Q\|_\infty \leq B$, $\|K\|_\infty \leq B$, and $\|V\|_\infty \leq B$. They derived their lower bound proof by building upon a different line of work that dealt with the fine-grained complexity of KDE problems, which was previously studied in \cite{bis17,acss20}. Their proof was based on a fine-grained reduction from the Approximate Nearest Neighbor search problem $\mathsf{ANN}$. Additionally, their findings explained how LLM computations can be made faster by assuming that matrix entries are bounded or can be well-approximated by a small number of bits, as previously discussed in \cite{zbiw19}, Section 2 and \cite{kvpf20}, Section 3.2.1. Specifically, they \cite{as23} showed a lower bound stating that when $B \geq \Omega(\sqrt{\log n})$, there is no algorithm that can approximate the computation in subquadratic time. However, when $B < o(\sqrt{ \log n})$, they proposed an algorithm that can approximate the attention computation almost linearly.

\paragraph{Transformer Theory}
Although the achievements of transformers in various fields are undeniable, there is still a significant gap in our precise comprehension of their learning mechanisms. Although these models have been examined on benchmarks incorporating numerous structured and reasoning activities, comprehending the mathematical aspects of transformers still considerably lags behind.
Prior studies have posited that the success of transformer-based models, such as BERT \cite{dclt18}, can be attributed to the information contained within its components, specifically the attention heads. These components have been found to hold a significant amount of information that can aid in solving various probing tasks related to syntax and semantics, as noted by empirical evidence found in several studies \cite{hm19,cklm19,tdp19,hl19,vb19,b22}.

Various recent studies have delved into the representational power of transformers and have attempted to provide substantial evidence to justify their expressive capabilities. 
These studies have employed both theoretical as well as controlled experimental methodologies through the lens of Turing completeness \cite{bpg20}, function approximation \cite{ybr+20}, formal language representation \cite{bag20, egz20, yppn21}, abstract algebraic operation learning \cite{zbb+22}, and statistical sample complexity \cite{wcm21, egkz22} aspects.  According to the research conducted by \cite{ybr+20}, transformers possess the capability of functioning as universal approximators for sequence-to-sequence operations. Similarly, the studies carried out by \cite{pmb19,bpg20} have demonstrated that attention models may effectively imitate Turing machines.
In addition to these recent works, there have been several previous studies that aimed to assess the capacity of neural network models by testing their learning abilities on simplistic data models \cite{ss92, yppn21, zbb+22}. 
Furthermore, \cite{llr23} conducted a formal analysis of the training dynamics to further understand the type of knowledge that the model learns from such data models.
According to findings from a recent study \cite{zpga23}, moderately sized masked language models have demonstrated the ability to parse with satisfactory results. Additionally, the study utilized BERT-like models that were pre-trained using the masked language modeling loss function on the synthetic text generated with probabilistic context-free grammar. The researchers empirically validated that these models can recognize syntactic information that aids in partially reconstructing a parse tree. \cite{lsz23} studied the computation of regularized version of exponential regression problem (without normalization factor).

\paragraph{Dynamic Maintenance}
In recent years, projection maintenance has emerged as a crucial data structure problem. The effectiveness and efficiency of several cutting-edge convex programming algorithms greatly hinge upon a sturdy and streamlined projection maintenance data structure~\cite{cls19,lsz19,b20,jlsw20,blss20,jswz21,sy21,dly21,b21,jkl+20,hjs+22,gs22}. 
There are two major differences between the problem in the dynamic data structure for optimization and our dynamic attention matrix maintenance problem. 
The first notable difference is that, in the optimization task, the inverse of a full rank square matrix is typically computed, whereas, in the attention problem, we care about the inverse of a positive diagonal matrix which behaves the normalization role in LLMs. 
The second major difference is, in the standard optimization task, all the matrix matrix operations are linear operations. However, in LLMs, non-linearity such as softmax/exp function is required to make the model achieve good performance. Therefore, we need to apply an entry-wise nonlinear function to the corresponding matrix. In particular, to compute $f(QK^\top)V$ when $f$ is linear function, we can pre-compute $K^\top V$. However when $f$ is $\exp$ function, we are not allowed to compute $K^\top V$ directly. 

Next, we will give more detailed reviews for classical optimization dynamic matrix maintenance problems. 
Let $B\in \R^{m\times n}$, consider the projection matrix $P=B^\top (BB^\top)^{-1}B$. The projection maintenance problem asks the following data structure problem: it can preprocess and compute an initial projection. At each iteration, $B$ receives a low rank or sparse change, and the data structure needs to update $B$ to reflect these changes. It will then be asked to approximately compute the matrix-vector product, between the updated $P$ and an online vector $h$. For example, in linear programming, one sets $B=\sqrt{W}A$, where $A\in \R^{m\times n}$ is the constraint matrix and $W$ is a diagonal matrix. In each iteration, $W$ receives relatively small perturbations. Then, the data structure needs to output an approximate vector to $\sqrt{W}A^\top (AWA^\top)^{-1} A\sqrt{W}h$, for an online vector $h\in \R^n$. 

\paragraph{Roadmap}
The rest of the paper is organized as follows.
In Section \ref{sec:pre}, we give some preliminaries. 
In Section \ref{sec:tech}, we explain the techniques used to show our upper bound and lower bound results.
In Section \ref{sec:mainupperbound}, we present our dynamic data-structure. Our algorithm shows the upper bound results.
In Section \ref{sec:mainlowerbound}, we give our conditional lower bound result by assuming the Hinted MV conjecture.
 %%% Section 1. Introduction

\section{Preliminary}
\label{sec:pre}

For a matrix $A$, we use $A^\top$ to denote its transpose. For a non-zero diagonal matrix $D \in \R^{n \times n}$, we use $D^{-1} \in \R^{n \times n}$ to denote the matrix where the $(i,i)$-th diagonal entry is $(D_{i,i})^{-1}$ for all $i \in [n]$.

For a vector $x \in \R^n$, we use $\diag(x) \in \R^{n \times n}$ to denote an $n \times n$ matrix where the $i,i$-th entry on the diagonal is $x_i$ and zero everywhere else for all $i \in [n]$.

In many TCS/ML literature, $\exp(M)$ denotes the matrix exponential, i.e., $\exp(M) = \sum_{i=0}^{\infty} \frac{1}{i!} M^i$. However, in this paper, we use $\exp(M)$ to denote the entry-wise exponential, i.e.,
\begin{align*}
\exp(M)_{i,j} := \exp( M_{i,j} ).
\end{align*}

We use ${\bf 1}_n$ to denote the length-$n$ vector where all the entries are ones. We use ${\bf 0}_n$ to denote the length-$n$ vector where all entries are zeros.

We give a standard fact that is used in our proof.
\begin{fact}[folklore]\label{fac:simple_matrix_stack}
Given a set of vectors $a_1, \cdots, a_k \in \R^n$ and $b_1, \cdots b_k \in \R^d$, then we have
\begin{align*}
\sum_{i=1}^k a_i b_i^\top = A B^\top
\end{align*}
where $A \in \R^{n \times k}$ and $a_i$ is $i$-th column of $A$, and $B \in \R^{d \times k}$ and $b_i$ is the $i$-th column of $B$ for all $i \in [k]$. 

Further, we have
\begin{itemize}
    \item Part 1. Computing $AB^\top$ 
    \begin{itemize}
        \item takes $O(nkd)$ time, if we do it naively 
        \item takes $\Tmat(n,k,d)$ time, if we use fast matrix multiplication
    \end{itemize}
    \item Part 2. For any matrix $C \in \R^{d \times d}$, computing $A B^\top C$ 
    \begin{itemize}
        \item takes $\Tmat(n,k,d) + \Tmat(n,d,d)$, if we use fast matrix multiplication, first compute $A B^\top$ then compute $(A B^\top) C$
        \item takes $\Tmat(k,d,d) + \Tmat(n,k,d)$ time, if we use fast matrix multiplication, first compute $B^\top C$, then compute $A(B^\top C)$
    \end{itemize}
\end{itemize}
\end{fact}

We define a standard notation for describing the running time of matrix multiplication, see literature \cite{di00,z02,s04,s05,lg14,bn19,cls19,lsz19,bns19,b20,gr21,jswz21,b21} for examples. 

\begin{definition}
For any three positive integers, we use $\Tmat(a,b,c)$ to denote the time of multiplying an $a \times b$ matrix with another $b \times c$ matrix.
\end{definition}
We use $\omega$ to denote the time that $n^{\omega} = \Tmat(n,n,n)$. Currently $\omega \approx 2.373$ \cite{w12,lg14,aw21}.

\begin{definition}
We define $\omega(\cdot, \cdot, \cdot)$ function as follows, for any $a,b$ and $c$, we use $\omega(a,b,c)$ to denote that $n^{\omega(a,b,c)} = \Tmat(n^a,n^b,n^c)$. 
\end{definition}

\section{Technique Overview}
\label{sec:tech}

Given three matrices $Q, K, V\in \R^{n \times d}$, we need to compute the attention given by
$
\Att(Q,K,V) = D^{-1} A V
$
where square matrix $A \in \R^{n \times n}$ and diagonal matrix $D \in \R^{n \times n}$ are $
A:= \exp(Q K^\top)$, $D := \diag( A {\bf 1}_n )
$. The static problem \cite{as23} is just computing $\mathsf{Att}$ for given $Q,K$ and $V$. In the dynamic problem, we can get updates for $K$ and $V$ in each iteration.

For the algorithmic result in \cite{as23}, they make use of the ``polynomial method in algorithm design''. The polynomial method is a technique for finding low-rank approximations of the attention matrix $A$, which can be computed efficiently if the entries are bounded. 
For the hardness result in \cite{as23}, they assume the strong exponential time hypothesis and use nearest neighbor search hardness result in the reduction.

\subsection{Algorithm}

\paragraph{Problem Formulation}

For each update, we receive $\delta$ as input and update one entry in either matrix $K$ or $V$. In the query function, we take index $i\in [n], j \in [d]$ as input, and return the $\{i,j\}$-th element in the target matrix $B:= D^{-1} A V$.

Let $C$ denote $A V$. Let $\wt{B}$ denote the updated target matrix $B$.  We notice that the computation of the attention can be written as 
\begin{align*}
    \wt{B} =  (D^{-1} + \Delta_D) (C + \Delta_C)
\end{align*}
Let $\Delta^{(t)}$ denote the change in the $t$-th iteration. 
In a lazy-update fashion, we write $\wt{B}$ in the implicit form 
\begin{align*}
      \wt{B} = (D^{-1} + \sum_{t=1}^{\mathrm{ct}}\Delta_D^{(t)}) (C + \sum_{t=1}^{\mathrm{ct}}\Delta_C^{(t)})
\end{align*}
where $\mathrm{ct}$ denotes the number of updates since the last time we recomputed $D$ and $C$.

\paragraph{Lazy Update}

We propose a lazy-update algorithm (Algorithm \ref{alg:fast_update_K}) that does not compute the attention matrix when there is an update on the key matrix $K$.
We also propose a lazy-update algorithm (Algorithm \ref{alg:fast_update_V}) that does not compute the attention matrix when there is an update  on the value matrix $V$.
Instead, we maintain a data-structure (Algorithm \ref{alg:fast}) that uses $\mathrm{List}_C, \mathrm{List}_D$ and $\mathrm{List}_V$ to record the update by storing rank-1 matrices before the iteration count reaches the threshold $n^a$ for some constant $a$. 
For the initialization (Algorithm \ref{alg:fast}), we compute the exact target matrix $D^{-1} A V$ and other intermediate matrices, which takes $O( \Tmat(n,d,n))$ time (Lemma \ref{lem:init}).

\paragraph{Re-compute}
When the iteration count reaches the threshold $n^a$, we re-compute all the variables in the data-structure as follows (Lemma \ref{lem:fast_recompute_correct}). 
By using Fact \ref{fac:simple_matrix_stack},  we first stack all the rank-$1$ matrices in $\mathrm{List}_C$ and compute the matrix multiplication once to get $\sum_{t=1}^{\mathrm{ct}}\Delta_C^{(t)}$ using $\Tmat(n,n^a,d) = n^{\omega(1,1,a)}$ time (Lemma \ref{lem:fast_recompute_time}). Then, we compute $C + \sum_{t=1}^{\mathrm{ct}}\Delta_C^{(t)}$ to get the re-computed $\wt{C}$. 
Similarly, to re-compute $V$, we stack all the rank-$1$ matrices in $\mathrm{List}_V$ and compute the matrix multiplication once to get $\sum_{t=1}^{\mathrm{ct}}\Delta_V^{(t)}$ using $\Tmat(n,n^a,d) = n^{\omega(1,1,a)}$ time. Then, we compute $V + \sum_{t=1}^{\mathrm{ct}}\Delta_V^{(t)}$ to get the re-computed $\wt{V}$. 
To re-compute the diagonal matrix $D$, we sum up all the updates by $ \sum_{t=1}^{\mathrm{ct}} \Delta_D^{(t)} $  and add it to the old $D^{-1}$ (detail can be found in Algorithm \ref{alg:fast_recompute}). Hence, our algorithm takes $n^{\omega(1,1,a)} / n^{a}$ amortized time to update $K$ and $V$ (Lemma \ref{lem:fast_update_K_time}, Lemma \ref{lem:fast_update_V_time}).

\paragraph{Fast Query}
Recall that the query function  takes index $i\in [n], j \in [d]$ as input, and returns the $\{i,j\}$-th element in the target matrix $B:= D^{-1} A V$. Let $\wt{D}^{-1}$ denote the lates $D^{-1}$ obtained from $\mathrm{List}_D$. Let $\Delta_{V,1}$ and $\Delta_{V,2}$ be stacked matrix obtained from list from $V$. We can rewrite the output by 
\begin{align*}
 & ~ ( (\wt{D}^{-1} ) \cdot ( A ) \cdot (V + \Delta_{V,1} \Delta_{V,2}) )_{i,j} \\
 =  & ~  ( (\wt{D}^{-1} ) \cdot ( A \cdot V  ))_{i,j} + ( (\wt{D}^{-1} ) \cdot A \cdot (\Delta_{V,1} \Delta_{V,2}) )_{i,j} \\
 = & ~  (\wt{D})_i^{-1} C_{i,j} + (\wt{D})_i^{-1} A_{i,*} \Delta_{V,1} (\Delta_{V,2})_{*,j}.
\end{align*}
Note that we maintain $C$ in our re-compute function. Hence, computing the first part takes $O(1)$ time. As each column of $\Delta_{V,1}$ and row of $\Delta_{V,2}$ is 1-sparse, computing the second part takes $O(n^a)$ time. The total running time needed for the query function is $O(n^a)$ (Lemma \ref{lem:fast_query_time}, Lemma \ref{lem:fast_query_correct}).

\subsection{Hardness}

We now turn to our lower bound result, which is inspired by the $\mathsf{HMV}$ conjecture \cite{bns19}. Let us firstly define the $\mathsf{HMV}$ problem (see formal definition in Definition~\ref{def:hintedMv}).

Let the computation be performed over the boolean semi-ring and let $m = n^{\tau}, \forall 0 < \tau \leq 1$. The $\mathsf{HMV}$ problem has the following three phases
\begin{itemize}
    \item {\bf Phase 1.} Input two $n \times n$ matrices $M$ and $V$
    \item {\bf Phase 2.} Input an $n\times n$ matrix $P$ with at most $n^\tau$ non-zero entries 
    \item {\bf Phase 3.} Input a single index $i \in [n]$ 
    \begin{itemize}
        \item We need to answer $M P V_{*,i}$
        \item Here $V_{*,i} \in \R^n$ is the $i$-th column of matrix $V$
    \end{itemize}
\end{itemize}

According to \cite{bns19}, the above problem is conjectured to be hard in the following sense,
\begin{conjecture}[Hinted MV ($\mathsf{HMV}$), \cite{bns19}]\label{con:hintedMv_informal}
For every constant $0 < \tau \leq 1$ no algorithm for the hinted Mv problem (Definition~\ref{def:hintedMv}) can simultaneously satisfy
\begin{itemize}
\item polynomial time in {\bf Phase 1.}
\item $O (n^{\omega(1,1, \tau)-\epsilon} )$ time complexity in {\bf Phase 2.} and
\item $O (n^{1+\tau-\epsilon} )$ in {\bf Phase 3.}
\end{itemize}
for some constant $\epsilon>0$.
\end{conjecture}

Our primary contribution lies in demonstrating how to reduce the $\mathsf{OAMV}$ (Definition \ref{def:AHAMV}) and $\mathsf{ODAMV}$ (Definition \ref{def:AHDAMV}) to the $\mathsf{HMV}$ problem (Definition \ref{def:hintedMv}). To achieve this, we have adopted a contradiction-based approach. Essentially, we begin by assuming the existence of an algorithm that can solve the $\mathsf{OAMV}$ problem with polynomial initialization time and amortized update time of $O(\Tmat(n, n^\tau, d) / n^{\tau + \Omega(1)})$, while worst-case query time is $O(n^{\tau - \Omega(1)} )$ for all $\tau \in (0,1]$. Our assumption implies that there exists a data structure that is faster than our result (Theorem \ref{thm:fast}).
We subsequently proceed to demonstrate that using this algorithm enables us to solve the $\mathsf{HMV}$ problem too quickly, which contradicts the $\mathsf{HMV}$ conjecture. 

Specifically, let us take an instance for the $\mathsf{HMV}$ problem (Definition~\ref{def:hintedMv})
\begin{itemize}
    \item Let $\mathsf{M},\mathsf{V} \in \{0,1\}^{n \times n}$ denote two matrices from {\bf Phase 1.} from $\mathsf{HMV}$.
\end{itemize}

We create a new instance $\mathsf{OAMV}(\wt{n} = n,\wt{d}=n)$ where 
\begin{align*}
\wt{Q} = \mathsf{M} , ~~~ \wt{K} = 0, ~~~ \wt{V} = \mathsf{V}
\end{align*}

In Claim~\ref{cla:1_case_AHAMV} and Claim~\ref{cla:0_case_AHAMV}, by making use of our construction of $\wt{Q}, \wt{K}$ and $\wt{V}$, we show that for each $i \in [n]$ and $j \in [n]$, 
 
\begin{align*}
\text{If}~~(( \exp(\wt{Q} \wt{K}^\top) - {\bf 1}_{n \times n} ) \wt{V})_{j,i}>&~ 0,\text{ then } (\mathsf{M} \mathsf{P} \mathsf{V})_{j,i}  = 1. \\
\text{If}~~((\exp(\wt{Q} \wt{K}^\top) - \mathbf{1}_{n \times n}) \wt{V})_{j,i} =&~ 0,\text{ then } (\mathsf{M} \mathsf{P} \mathsf{V})_{j,i}  = 0.
\end{align*}
 
By using the above two statements, we know that $\exp(\wt{Q}\wt{K}^\top)\wt{V}_{*,i}$ is enough to reconstruct $\mathsf{M}\mathsf{P}\mathsf{V}_{*,i}$ for the $\mathsf{HMV}$ problem (Definition~\ref{def:hintedMv}). Then, solving $\mathsf{M}\mathsf{P}\mathsf{V}_{*,i}$ takes polynomial initialization time and amortized update time of $O(\Tmat(n, n^\tau, d) / n^{\tau + \Omega(1)})$, while worst-case query time is $O(n^{\tau - \Omega(1)} )$ for every $\tau \in (0,1]$. The contradiction of the $\mathsf{HMV}$ conjecture shows that there is no such algorithm.
Similarly, for the normalized case $\mathsf{ODAMV}$ (Definition \ref{def:AHDAMV}) problem, we show how to reconstruct another instance of the $\mathsf{HMV}$ problem and complete the proof by contradiction.

\section{Main Upper Bound}\label{sec:mainupperbound}

In Section \ref{sec:upper_init}, we show the running time of initializing our data structure.
In Section \ref{sec:upper_update}, we show the running time of updating $K$ and $V$.
In Section \ref{sec:upper_query}, we show the correctness and the running time of querying the target matrix.
In Section \ref{sec:upper_recompute}, we show the correctness and the running time of recomputing the variables in our data-structure.

We propose our upper bound result as the following: 
\begin{theorem}[Main algorithm, formal version of Theorem~\ref{thm:fast_informal}]\label{thm:fast}
For any constant $a \in (0,1]$. Let $d = O(n)$.
There is a dynamic data structure that uses $O(n^2 )$ 
space and supports the following operations:
\begin{itemize}
    \item \textsc{Init}$(Q,K,V) $. It runs in $O( \Tmat(n,d,n))$ time.
    \item \textsc{UpdateK}$(i \in [n], j \in [d], \delta \in \R)$. This operation updates one entry in $Q$, and it runs in $O(\Tmat(n,n^a, n) / n^a )$ amortized time.
    \item \textsc{UpdateV}$(i \in [n], j \in [d], \delta \in \R)$. This operation takes same amortized time as $Q$ update.
    \item \textsc{Query}$(i \in [n],j \in [d])$. This operation outputs $(D^{-1} (\exp(QK^\top)) V)_{i,j}$ operation takes in $O(n^a)$ worst case time.
\end{itemize}
\end{theorem}
\begin{remark}
The amortized time in \textsc{UpdateK} and \textsc{UpdateV} can be made into worst case time by using standard techniques, e.g. see Section B of \cite{bns19}. 
\end{remark}

\begin{algorithm}[!ht]\caption{Dynamic Data Structure }\label{alg:fast}
\begin{algorithmic}[1]
\State {\bf data structure} \textsc{DynamicAttention} \Comment{Theorem~\ref{thm:fast}}
\State {\bf members}
\State \hspace{4mm} $Q \in \R^{n \times d}$ \Comment{Query token}
\State \hspace{4mm} $K \in \R^{n \times d}$ \Comment{Key token}
\State \hspace{4mm} $V \in \R^{n \times d}$ \Comment{Value token}
\State \hspace{4mm} $M \in \R^{n \times n}$ \Comment{The logits matrix, $M = Q K^\top$}
\State \hspace{4mm} $A \in \R^{n \times n}$ \Comment{The attention matrix, $A = \exp(QK^\top)$}
\State \hspace{4mm} $D \in \R^{n \times n}$ \Comment{The diagonal matrix, }
\State \hspace{4mm} $C \in \R^{n \times d}$ \Comment{Intermediate matrix, $C = \exp(QK^\top) V$}
\State \hspace{4mm} $B \in \R^{n \times d}$ \Comment{Target matrix, $B=D^{-1}A V$}
\State \hspace{4mm} $\mathrm{List}_A$ \Comment{List with size $n^a$}
\State \hspace{4mm} $\mathrm{List}_C$ \Comment{List with size $n^a$}
\State \hspace{4mm} $\mathrm{List}_D$ \Comment{List with size $n^a$}
\State \hspace{4mm} $\mathrm{ct}_K, \ct_V$
\State {\bf end members}
\State
\Procedure{Init}{$Q,K, V$} \Comment{Lemma~\ref{lem:init}}
    \State $Q \gets Q$, $K \gets K$, $V \gets V$
    \State $M \gets QK^\top$, $A \gets \exp(QK^\top)$
    \State $C \gets \exp(QK^\top) V$
    \State $B \gets D^{-1} A V$
    \State $\ct_K \gets 0$
    \State $\ct_V \gets 0$
\EndProcedure
\State {\bf end data structure}
\end{algorithmic}
\end{algorithm}

\begin{algorithm}[!ht]\caption{Algorithm that update $K$ and maintain the data structure  }\label{alg:fast_update_K}
\begin{algorithmic}[1]
\State {\bf data structure} \textsc{DynamicAttention} \Comment{Theorem~\ref{thm:fast}}
\Procedure{UpdateK}{$i \in [n], j \in [d], \delta$} \Comment{Lemma~\ref{lem:fast_update_K_time}}
    \State $\mathrm{ct}_K \gets \mathrm{ct}_K + 1 $
    \State $\wt{K}_{i,j} \gets K_{i,j} + \delta$
    \State $(\Delta_M)_{*,i} \gets \delta  \cdot \underbrace{ Q }_{n \times d}   \underbrace{ e_j }_{d \times 1}$ \Comment{$\Delta_M $ only have entries in $i$-th column}
   
    \State \Comment{Here $\circ$ denotes entry-wise product}
    \State $(\Delta_{A})_{*,i} \gets ( A_{*,i}\circ (\exp ((\Delta_M)_{*,i}) - {\bf 1}_n) )$
     \State  $\wt{M} \gets M + (\Delta_M)_{*,i} e_i^\top$ \Comment{We only update $i$-th column of $M$}
    \State  $\wt{A} \gets A + (\Delta_A)_{*, i} e_i^\top $ \Comment{We only update $i$-th column of $A$}
    
    \State Obtain diagonal vector $D_{\mathrm{tmp}}$ from $\mathrm{List}_D[\ct_K - 1].\textsc{Getb}$ \Comment{It takes $O(n)$ time}
    \State $\wt{D} \gets D^{-1}_{\mathrm{tmp}} + \diag ( \Delta_A )_{*,i}$
    \For{$j=1 \to n$}
        
        \State  $(\Delta_D)_{j,j} \gets (D_{\mathrm{tmp}})_{j,j}^{-1} - \wt{D}_{j,j}^{-1}$ 
    \EndFor
    \If{$\mathrm{ct}_K < n^a$} 
        \State $\mathrm{List}_C[\mathrm{ct}_K-1 ].(a,b) \gets ( ( \Delta_A )_{*,i} \in \R^n, V^\top e_i \in \R^d)$
        \State $\mathrm{List}_D[\mathrm{ct}_K-1 ].(a,b) \gets (\Delta_D \in \R^{n \times n}, \wt{D}^{-1} \in \R^{n \times n})$ \Comment{Diagonal matrices}
    \Else  \Comment{$\Tmat(n,n^a,d) = n^{\omega(1,1,a)}$ time}
        \State \textsc{ReCompute}() \Comment{Algorithm \ref{alg:fast_recompute}. Re-compute everything}
       
    \EndIf
    \State {\color{blue}/*Referesh the memory*/}
    \State $K \gets \wt{K}$
    \State $A \gets \wt{A}$
    \State $M \gets \wt{M}$ 
\EndProcedure
\State {\bf end data structure}
\end{algorithmic}
\end{algorithm}

\begin{algorithm}[!ht]\caption{}\label{alg:fast_update_V}
\begin{algorithmic}[1]
\State {\bf data structure} \textsc{DynamicAttention} \Comment{Theorem~\ref{thm:fast}}
\Procedure{UpdateV}{$i \in [n], j \in [d], \delta$} \Comment{Lemma~\ref{lem:fast_update_V_time}}
    \State $\ct_V \gets \ct_V + 1$
    \If{$\ct_V < n^a$}
        \State $\mathrm{List}_V[ \ct_V - 1 ].(a,b) \gets (e_i \in \R^n, \delta e_j \in \R^d)$
    \Else
        \State $\textsc{ReCompute}()$ \Comment{Algorithm \ref{alg:fast_recompute}. Re-compute everything}
    \EndIf
\EndProcedure
\State {\bf end data structure}
\end{algorithmic}
\end{algorithm}

\begin{algorithm}[!ht]\caption{Algorithm that query the $\{i,j\}$-th element in the target matrix}\label{alg:fast_query}
\begin{algorithmic}[1]
\State {\bf data structure} \textsc{DynamicAttention} \Comment{Theorem~\ref{thm:fast}}
\Procedure{Query}{$i \in [n],j \in [d]$} \Comment{Lemma~\ref{lem:fast_query_time}, \ref{lem:fast_query_correct}}
    \State Let $\Delta_{V,1}$ and $\Delta_{V,2}$ be rectangular matrix obtained from list from $V$
    \State Let $(D_{\mathrm{tmp}})_i^{-1}$ denote the list of diagonal matrices obtained from $\mathrm{List}_D[\ct_K].\textsc{Getb}$ \Comment{This takes $O(1)$ time}
    \State {\color{blue}/*Below is the target*/}
    \State $\text{answer} \gets ( (D_{\mathrm{tmp}}^{-1} ) \cdot ( A ) \cdot (V + \Delta_{V,1} \Delta_{V,2}) )_{i,j} $ 
    \State {\color{blue}/*The actual computation*/}
    \State {\color{blue}/*Part 1. Answer, This is fast because we store $C=AV$*/}
    \State $\text{answer}_1 \gets  (D_{\mathrm{tmp}})_i^{-1} C_{i,j}$ \Comment{$O(1)$ time}
    \State {\color{blue}/*Part 2. Answer, this is fast because each column of $\Delta_{V,1}$ and row of $\Delta_{V,2}$ is 1-sparse*/}
    \State $\text{answer}_2 \gets  (D_{\mathrm{tmp}})_i^{-1} A_{i,*} \Delta_{V,1} (\Delta_{V,2})_{*,j} $ \Comment{$n^a$ time}
    \State $\text{answer} \gets \sum_{j=1}^2 \text{answer}_j$
    \State \Return \text{answer}
\EndProcedure
\State {\bf end data structure}
\end{algorithmic}
\end{algorithm}

\begin{algorithm}[!ht]\caption{Algorithm that re-compute evreything}\label{alg:fast_recompute}
\begin{algorithmic}[1]
\State {\bf data structure} \textsc{DynamicAttention} \Comment{Theorem~\ref{thm:fast}}
\Procedure{Recompute}{$ $} \Comment{Lemma~\ref{lem:fast_recompute_time}, Lemma~\ref{lem:fast_recompute_correct}}
    \State Let $\Delta_{C,1}$ and $\Delta_{C,2}$ be rectangular matrix obtained from $\mathrm{List}_C$
    \State Let $\Delta_{V,1}$ and $\Delta_{V,2}$ be rectangular matrix obtained from $\mathrm{List}_V$
    \State Let $\Delta_D(i)$ denote the list of diagonal matrices obtained from $\mathrm{List}_D[i].\textsc{Geta}$
    \State $\wt{C} \gets C + \Delta_{C,1} \cdot \Delta_{C,2}$ \Comment{It takes $\Tmat(n,n^a,d)$ time}
    \State $\wt{V} \gets V + \Delta_{V,1} \cdot \Delta_{V,2}$ \Comment{It takes $\Tmat(n,n^a,d)$ time}
    \State $\Delta_D \gets \sum_{i=1}^{\ct_K} \Delta_D(i) $ \Comment{it takes $n^{1+a}$ time}
    \State $\wt{D}^{-1} \gets  D^{-1} + \Delta_D$  \Comment{It takes $n$ time}
    
    \State $\wt{B} \gets \wt{D}^{-1} \cdot \wt{C} $ \Comment{This takes $nd$}   
    \State {\color{blue}/*Refresh the memory*/}
    \State $D\gets \wt{D}, C \gets \wt{C}, B \gets \wt{B}$, $V \gets \wt{V}$
    \State {\color{blue}/*Reset the counter*/}
    \State $\ct_K \gets 0$, $\ct_V \gets 0$ 
\EndProcedure
\State {\bf end data structure}
\end{algorithmic}
\end{algorithm}

\subsection{Initialization}\label{sec:upper_init}
We first give the running time of the initialization procedure.

\begin{lemma}[Init]\label{lem:init}
The procedure \textsc{Init} (Algorithm~\ref{alg:fast}) takes $\Tmat(n,d,n)$ time.
\end{lemma}
\begin{proof}
It is trivially from applying fast matrix multiplication.
\end{proof}

\subsection{Update}\label{sec:upper_update}

Next, we give the running time of updating $K$.
\begin{lemma}[Running time of \textsc{UpdateK}]\label{lem:fast_update_K_time}
The procedure \textsc{UpdateK} (Algorithm \ref{alg:fast_update_K}) takes
\begin{itemize}
    \item Part 1. $\Tmat(n,n,n^a)$  time in the worst case
    \item Part 2. $\Tmat(n,n,n^a)/ n^{a}$ time in the amortized case
\end{itemize}
\end{lemma}

\begin{proof}

{\bf Part 1.} It trivially from Lemma~\ref{lem:fast_recompute_time}

{\bf Part 2.} If the $\ct_K< n^a$, we pay $O(n)$ time. If $\ct_K = n^a$, we pay $n^{\omega(1,1,a)}$. So the amortized time is
\begin{align*}
\frac{ n (n^a-1)+ n^{\omega(1,1,a)} }{ n^a} = O(n^{\omega(1,1,a)-a})
\end{align*}
Note that, by using fast matrix multiplication and the fact that $d = O(n)$, we have 
$ n^{\omega(1,1,a)} = \Tmat(n,n^a,d)$.
Thus we complete the proof.
\end{proof}

Now, we give the running time of updating $V$.
\begin{lemma}[Running time of \textsc{UpdateV}]\label{lem:fast_update_V_time}
The procedure \textsc{UpdateV} (Algorithm \ref{alg:fast_update_V}) takes
\begin{itemize}
    \item Part 1. $\Tmat(n,n,n^a)$  time in the worst case.
    \item Part 2. $\Tmat(n,n,n^a)/ n^{a}$ time in the amortized case.
\end{itemize}
\end{lemma}

\begin{proof}

{\bf Part 1.} It trivially from Lemma~\ref{lem:fast_recompute_time}.

{\bf Part 2.} If the $\ct_K< n^a$, we pay $O(n)$ time. If $\ct_K = n^a$, we pay $n^{\omega(1,1,a)}$. So the amortized time is
\begin{align*}
\frac{ n (n^a-1)+ n^{\omega(1,1,a)} }{ n^a} = O(n^{\omega(1,1,a)-a})
\end{align*}

Note that, by using fast matrix multiplication and the fact that $d = O(n)$, we have 
$ n^{\omega(1,1,a)} = \Tmat(n,n^a,d)$.
Thus we complete the proof.
\end{proof}

\subsection{Query}\label{sec:upper_query}
We show the correctness of our \textsc{Query} that queries only one element in the target matrix.
\begin{lemma}[Correctness of \textsc{Query}] \label{lem:fast_query_correct}
The procedure \textsc{Query} (Algorithm \ref{alg:fast_query}) outputs 
\begin{align*}
\wt{B}_{i,j} = & ~ (D^{-1} \cdot A \cdot (V + \Delta_V))_{i,j} \\
= & ~ (D^{-1} A V +  D^{-1} A \Delta_V )_{i,j} 
\end{align*}
\end{lemma}
\begin{proof}
Let $\Delta_{V,1}$ denote the vector obtained from $\mathrm{List}_D[\ct_K].\textsc{Geta}$.

Let $\Delta_{V,2}$ denote the vector obtained from $\mathrm{List}_D[\ct_K].\textsc{Getb}$

Let $(D_{\mathrm{tmp}})_i^{-1}$ denote the list of diagonal matrices obtained from $\mathrm{List}_D[\ct_K].\textsc{Getb}$ 

We know
\begin{align*}
    \wt{B} =  &  ~ ( (D_{\mathrm{tmp}}^{-1} ) \cdot ( A ) \cdot (V + \Delta_{V,1} \Delta_{V,2}) ) \\
    =   &  ~ (D_{\mathrm{tmp}})^{-1} (AV) + (D_{\mathrm{tmp}})^{-1} ( A \Delta_{V,1} \Delta_{V,2})
\end{align*}

For the $\{i,j\}$-th element, by using simple algebra, we have
\begin{align*}
    \wt{B}_{i,j} = &  ~ (D_{\mathrm{tmp}})_i^{-1} (AV)_{i,j} + (D_{\mathrm{tmp}})_i^{-1} ( A \Delta_{V,1} \Delta_{V,2})\\
    = &  ~ (D_{\mathrm{tmp}})_i^{-1} (C)_{i,j} + (D_{\mathrm{tmp}})_i^{-1} ( A \Delta_{V,1} \Delta_{V,2})_{i,j} \\
    = &  ~ (D_{\mathrm{tmp}})_i^{-1} (C)_{i,j} + (D_{\mathrm{tmp}})_i^{-1} A_{i,*} \Delta_{V,1} (\Delta_{V,2})_{*,j}  \\
\end{align*}

We know 
\begin{align*}
    \text{answer}_1  = (D_{\mathrm{tmp}})_i^{-1} C_{i,j}
\end{align*}
and 
\begin{align*}
    \text{answer}_2 =  (D_{\mathrm{tmp}})_i^{-1} A_{i,*} \Delta_{V,1} (\Delta_{V,2})_{*,j} 
\end{align*}
By summing up $\text{answer}_1$ and $\text{answer}_2$, we have
\begin{align*}
    \wt{B}_{i,j}= & ~ (D^{-1} A V +  D^{-1} A \Delta_V )_{i,j} .
\end{align*}
Now, we complete the proof.
\end{proof}
Next, we give the running time of it.
\begin{lemma}[Running time of \textsc{Query}]\label{lem:fast_query_time} 
The running time of procedure \textsc{Query} (Algorithm \ref{alg:fast_query}) is $O(n^a)$.
\end{lemma}
\begin{proof}
We first stack all the vectors in $\mathrm{List}_V$ to $\Delta_{V,1} \in \R^{n \times n^a}$ and $\Delta_{V,2} \in \R^{n^a \times d}$, which takes $O(1)$ time.
\begin{itemize}
    \item Computing $ (D_{\mathrm{tmp}})_i^{-1} C_{i,j}$ takes $O(1)$ time.
    \item Computing $ (\Delta_{V,1} \Delta_{V,2})$ takes $O(n^a)$ time as $\Delta_{V,1}$ is $1$-sparse in columns and $(\Delta_{V,2})$ is $1$-sparse in rows.
    \item Computing $(D_{\mathrm{tmp}})_i^{-1} A_{i,*} (\Delta_{V,1} \Delta_{V,2})_{*,j}$ takes $O(n^a)$ time as $\nnz((\Delta_{V,1} \Delta_{V,2})_{*,j}) \leq n^a$.
\end{itemize}
Hence, the total running time needed is $O(n^a)$
\end{proof}

\subsection{Re-compute}\label{sec:upper_recompute}
We show the correctness of our re-compute function.
\begin{lemma}[Correctness of \textsc{Recompute}]\label{lem:fast_recompute_correct}
The procedure \textsc{Recompute} (Algorithm \ref{alg:fast_recompute}) correctly re-compute $D,C,B,V$.
\end{lemma}
\begin{proof}

{\bf Part 1.} Re-compute $D$

Let $\Delta_D(i)$ denote the list of diagonal matrices obtained from $\mathrm{List}_D[i].\textsc{Geta}$.
Then, the total difference between the updated  $\wt{D}$ and $D$ is $ \sum_{i=1}^{\ct_K} \Delta_D(i)$.

By computing  $\wt{D}^{-1} \gets  D^{-1} + \Delta_D$, we correctly get the updated $\wt{D}^{-1}$. By computing the inverse of a diagonal matrix we get $\wt{D}$.

{\bf Part 2.} Re-compute $V$

We first stack all the vectors in $\mathrm{List}_V$ to $\Delta_{V,1} \in \R^{n \times n^a}$ and $\Delta_{V,2} \in \R^{n^a \times d}$.

By using Fact \ref{fac:simple_matrix_stack}, we have $\wt{V} = V + \Delta_{V,1} \cdot \Delta_{V,2}$.

{\bf Part 3.} Re-compute $C$

Similar to the proof of re-computing $V$.

We first stack all the vectors in $\mathrm{List}_C$ to $\Delta_{C,1} \in \R^{n \times n^a}$ and $\Delta_{C,2} \in \R^{n^a \times d}$.

By using Fact \ref{fac:simple_matrix_stack}, we have $\wt{C} = C + \Delta_{C,1} \cdot \Delta_{C,2}$.

{\bf Part 4.} Re-compute $B$

By using the definition of $B = D^{-1} C$, we can update $B$ by using $\wt{B} = \wt{D}^{-1} \cdot \wt{C}$.

Now, we complete the proof.
\end{proof}

Next, we give the running time of it.
\begin{lemma}[Running time of \textsc{Recompute}]\label{lem:fast_recompute_time}
The running time of procedure \textsc{Recompute} (Algorithm \ref{alg:fast_recompute}) is $\Tmat(n,n^a,d)$.
\end{lemma}
\begin{proof}
We first stack all the vectors in $\mathrm{List}_V$ to $\Delta_{V,1} \in \R^{n \times n^a}$ and $\Delta_{V,2} \in \R^{n^a \times d}$, which takes $O(1)$ time.

We  stack all the vectors in $\mathrm{List}_C$ to $\Delta_{C,1} \in \R^{n \times n^a}$ and $\Delta_{C,2} \in \R^{n^a \times d}$, which takes $O(1)$ time.

\begin{itemize}
    \item Computing $ C + \Delta_{C,1} \cdot \Delta_{C,2}$ takes $\Tmat(n,n^a,d)$ time.
    \item Computing $ V + \Delta_{V,1} \cdot \Delta_{V,2}$ takes $\Tmat(n,n^a,d)$ time.
    \item Computing $ \sum_{i=1}^{\ct_K} \Delta_D(i) $ takes $O(n^{a+1})$ time as $\nnz(\Delta_D(i) ) = O(n)$ and ${\ct_K} = O(n^a)$.
    \item Computing $ D^{-1} + \Delta_D$ takes $O(n)$ time as $\nnz(\Delta_D ) = O(n)$.
    \item Computing $ \wt{D}^{-1} \cdot \wt{C} $ takes $O(nd)$ time as $\wt{D}^{-1}$ is a diagonal matrix.
Hence, the total running time is $\Tmat(n,n^a,d)$.
\end{itemize}
\end{proof}

\section{Main Lower Bound}\label{sec:mainlowerbound}
In Section \ref{sec:lower_omv}, we give the definition of Online Matrix Vector ($\mathsf{OMV}$) problem.
In Section \ref{sec:lower_hardness}, we introduce the definition of Hinted MV and its conjecture (from previous work \cite{bns19}).
In Section \ref{sec:lower_notnormalized}, we show the hardness of computing the target matrix without the normalization factor.
In Section \ref{sec:lower_normalized}, we show the hardness of computing the target matrix with the normalization factor.
\subsection{Online Matrix Vector Multiplication}\label{sec:lower_omv}
Before studying the hardness of our problem, we first review a famous problem in theoretical computer science which is called  online matrix vector multiplication problem. 
Here is the definition of online matrix vector multiplication, which has been a crucial task in many fundamental optimization problems.

\begin{definition}[Online Matrix Vector ($\mathsf{OMV}$) \cite{hkns15,lw17, ckl18}]
Given a matrix $A \in \{0,1\}^{n \times n}$, let $T=O(n)$, there is an online sequence of vectors $u_1, \cdots, u_T \in \{0,1\}^n$. The goal is to design a structure that whenever receives a new vector $u_t$ and output $A u_t$.
\end{definition}

Such a problem is widely believed in the community that there is no algorithm to solve it in truly subquadratic time per vector and there is no algorithm to solve it in truly subcubic time over all vectors.

\subsection{Hardness from Previous Work}\label{sec:lower_hardness}

We define the hinted Mv problem from previous work \cite{bns19}.
\begin{definition}[Hinted MV ($\mathsf{HMV}$) \cite{bns19}]\label{def:hintedMv}
    Let the computations be performed over the boolean semi-ring and let $m = n^\tau$, $0 < \tau \leq 1$. The hinted $Mv$ problem consists of the following phases:
    \begin{enumerate}
        \item Input two $n \times n$ matrices $M$ and $V$ \label{phase1}
        \item Input an $n\times n$ matrix $P$ with at most $n^\tau$ non-zero entries \label{phase2}
        \item Input a single index $i \in [n]$ \label{phase3}
        \begin{itemize}
            \item We need to answer $M P V_{*,i}$
            \item Here $V_{*,i} \in \R^n$ is the $i$-th column of matrix $V$
        \end{itemize}  
    \end{enumerate}
\end{definition}

We give the hinted Mv conjecture which is from prior work \cite{bns19}.
\begin{conjecture}[$\mathsf{HMV}$ conjecture \cite{bns19}, restatement of Conjecture~\ref{con:hintedMv_informal}]\label{con:hintedMv}
For every constant $0 < \tau \leq 1$ no algorithm for the hinted Mv problem (\Cref{def:hintedMv}) can simultaneously satisfy
\begin{itemize}
\item polynomial time in phase \ref{phase1}
\item $O (n^{\omega(1,1, \tau)-\epsilon} )$ time complexity in phase \ref{phase2} and
\item $O (n^{1+\tau-\epsilon} )$ in phase \ref{phase3}
\end{itemize}
for some constant $\epsilon>0$.
\end{conjecture}

\subsection{Online Attention Matrix Vector Multiplication}\label{sec:lower_notnormalized}

We define the dynamic attention matrix vector problem here. For the following definition, we ignore the effect by the normalization factor. We will handle it in the later section.
\begin{definition}[$\mathsf{OAMV}(n,d)$]\label{def:AHAMV}
The goal of the \textbf{\textit{O}}nline \textbf{\textit{A}}ttention \textbf{\textit{M}}atrix \textbf{\textit{V}}ector Multiplication problem $\mathsf{OAMV}(n,d)$ is to design a data structure that satisfies the following operations:
\begin{enumerate}
    \item \textsc{Init}: Initialize on $n \times d$ matrices $Q$, $K$, $V$.
    \item \textsc{Update}: Change any entry of $Q$, $K$, or $V$.
    \item \textsc{Query}: For any given $i \in [n]$, $j\in[d]$, return $(\exp(QK^\top)V)_{i,j}$ .
\end{enumerate}
\end{definition}

Next, we present our lower bound result ignoring the normalization factor.
\begin{lemma}\label{lem:lowerbound}
Assuming the hinted $Mv$ conjecture (\Cref{con:hintedMv}): For every constant $0 < \tau \le 1$, there is no dynamic algorithm for $\mathsf{OAMV}(n,d)$ problem (Definition~\ref{def:AHAMV}) with
\begin{itemize}
    \item polynomial initialization time, and
    \item amortized update time $O(\Tmat(n, n^\tau, d) / n^{\tau + \Omega(1)})$, and
    \item worst query time $O(n^{\tau - \Omega(1)} )$.
\end{itemize}
\end{lemma}
\begin{proof}
Assume there was a dynamic algorithm faster than what is stated in \Cref{lem:lowerbound} for some parameter $\tau$, i.e.~update time $O(\Tmat(n, n^\tau, d) / n^{\tau + \epsilon})$ and query time $O(n^{\tau - \epsilon} )$ for some constant $\epsilon > 0$. We show that this would contradict the hinted $Mv$ conjecture (\Cref{con:hintedMv}).

Let us take an instance for the $v$-hinted Mv problem (\Cref{def:hintedMv}) 
with $\mathsf{M},\mathsf{V} \in \{0,1\}^{n \times n}$.
We create a new instance $\mathsf{OAMV}(\wt{n} = n,\wt{d}=n)$ where 
\begin{align*}
\wt{Q} = \mathsf{M} , ~~~ \wt{K} = 0, ~~~ \wt{V} = \mathsf{V}
\end{align*}
During phase \ref{phase1}, we give this input to the dynamic algorithm for the $\mathsf{OAMV}$ problem (Definition~\ref{def:AHAMV}).
During phase \ref{phase2}, when we receive the $n\times n$ matrix $\mathsf{P}$ with $n^\tau$ non-zero entries, we perform $n^\tau$ updates to the data structure to set $\wt{K}^\top = \mathsf{P}$. This takes
\begin{align*}
O(\wt{n}^\tau \cdot (\Tmat(\wt{n}, \wt{n}^\tau, \wt{d}) / \wt{n}^{\tau + \epsilon})) = O(n^{\omega(1,1,\tau)-\epsilon})
\end{align*}
time.

At last, in phase \ref{phase3}, we perform $\wt{n}$ queries to obtain the column $\exp(\wt{Q}\wt{K}^\top)\wt{V}_{*,i}$ in $O(\wt{n} \cdot \wt{n}^{\tau-\epsilon}) = O(n^{1+\tau-\epsilon})$ time.

Using Claim~\ref{cla:1_case_AHAMV}, and Claim~\ref{cla:0_case_AHAMV}, 
 we know that $\exp(\wt{Q}\wt{K}^\top)\wt{V}_{*,i}$ is enough to reconstruct $\mathsf{M}\mathsf{P}\mathsf{V}_{*,i}$ for the hinted $Mv$ problem.

\end{proof}

\begin{claim}\label{cla:1_case_AHAMV}
For each $i \in [n]$ and $j \in [n]$, if  $(( \exp(\wt{Q} \wt{K}^\top) - {\bf 1}_{n \times n} ) \wt{V})_{j,i}$ is  $> 0$, then $ (\mathsf{M} \mathsf{P} \mathsf{V})_{j,i}  = 1$,
\end{claim}
\begin{proof}
Assume we have
 \begin{align*}
    ( (\exp(\wt{Q} \wt{K}^\top) - {\bf 1}_{n \times n} ) \wt{V})_{j,i}  >   & ~  0,
 \end{align*}
We defined $\wt{Q} = \mathsf{M} , \wt{K} = \mathsf{P},  \wt{V} = \mathsf{V}$, so we can rewrite it as
 \begin{align*}
      ( (\exp(\mathsf{M}\mathsf{P}) - {\bf 1}_{n \times n} ) \mathsf{V})_{j,i}  >  & ~  0.
 \end{align*}
Using the definition of matrix multiplication, and the fact that $\exp(x)>1$ for all $x>0$, we have some $k \in [n]$ with
\begin{align*}
      ( (\exp(\mathsf{M}\mathsf{P}) - {\bf 1}_{n \times n} )_{j,k} (\mathsf{V})_{k,i}  >   & ~ 0 \\
      ( (\exp(\mathsf{M}\mathsf{P})_{j,k} - 1) (\mathsf{V})_{k,i}  >  & ~  0 
 \end{align*}
We can conclude that for each $i \in [n], j \in [n]$, there is at least one $k \in [n]$ such that 
\begin{itemize}
    \item $\mathsf{V}_{k,i} > 0$
    \item $(\mathsf{M}\mathsf{P})_{j,k} > 0$
\end{itemize}
Therefore, by using the definition of boolean semi-ring, we can conclude that $(\mathsf{M}\mathsf{P} \mathsf{V})_{j,i}  = 1$
 
\end{proof}

\begin{claim}\label{cla:0_case_AHAMV}
For each $i \in [n]$ and $j \in [n]$, if $((\exp(\wt{Q} \wt{K}^\top) - \mathbf{1}_{n \times n}) \wt{V})_{j,i}$ is $0$ then $(\mathsf{M} \mathsf{P} \mathsf{V})_{j,i}  = 0$.
\end{claim}
\begin{proof}

We have 
 \begin{align*}
    & ~ ( (\exp(\wt{Q} \wt{K}^\top) - \mathbf{1}_{n \times n})\wt{V})_{j,k} \\
   = & ~( (\exp(\wt{Q} \wt{K}^\top) - \mathbf{1}_{n \times n}))_{j,*} \wt{V}_{*,i } \\
   = & ~( (\exp(\mathsf{M}\mathsf{P}) - \mathbf{1}_{n \times n}))_{j,*} \mathsf{V}_{*,i } 
 \end{align*}
where the first step follows from the definition of matrix multiplication and the second step follows from the definition of $\wt{Q}, \wt{K}$ and $\wt{V}$.

By using the above equation, if $ ((\exp(\wt{Q} \wt{K}^\top) - \mathbf{1}_{n \times n})\wt{V})_{j,k} = 0$, we have
\begin{align}\label{eq:exp_1_v_0_AHAMV}
     (\exp(\mathsf{M} \mathsf{P} ) - \mathbf{1}_{n \times n})_{j,*} \mathsf{V}_{*,i }  =   0
\end{align}
Eq.~\eqref{eq:exp_1_v_0_AHAMV} implies that, for all $k \in [n]$ such that $\mathsf{V}_{k,i }  =  1$ , we have $(\exp(\mathsf{M} \mathsf{P}) - \mathbf{1}_{n \times n})_{j,k} = 0$ , which also implies that $(\mathsf{M}\mathsf{P})_{j,k} = 0$.

 Now, we can conclude that  $(\mathsf{M} \mathsf{P} \mathsf{V})_{j,i}  = 0$ for each $i \in [n]$ and $j \in [n]$.
\end{proof}

\subsection{Online Diagonal-normalized Attention Matrix Vector Multiplication}\label{sec:lower_normalized}

Next, we consider the normalization factor and defined the problem as the following.
\begin{definition}[$\mathsf{ODAMV}(n,d)$, restatement of Definition~\ref{def:AHDAMV_informal}]\label{def:AHDAMV}

The goal of \textbf{\textit{O}}nline \textbf{\textit{D}}iagonal-based normalized \textbf{\textit{A}}ttention \textbf{\textit{M}}atrix \textbf{\textit{V}}ector Multiplication problem $\mathsf{ODAMV}(n,d)$ is to design a data structure that satisfies the following operations:
\begin{enumerate}
    \item \textsc{Init}: Initialize on $n \times d$ matrices $Q$, $K$, $V$.
    \item \textsc{Update}: Change any entry of $Q$, $K$, or $V$.
    \item \textsc{Query}: For any given $i \in [n]$, $j\in[d]$, return $(D^{-1}\exp(QK^\top)V)_{i,j}$, where  $D = \diag( \exp(QK^\top) {\bf 1}_n )$.
\end{enumerate}
\end{definition}

Next, we present our lower bound result with the normalization factor.
\begin{lemma}\label{lem:lowerbound_D}

Assuming the hinted $Mv$ conjecture (\Cref{con:hintedMv}): For every constant $0 < \tau \le 1$, there is no algorithm that solve $\mathsf{ODAMV}(n,d)$ problem (Definition~\ref{def:AHDAMV}) with 
\begin{itemize}
    \item polynomial initialization time, and
    \item amortized update time $O(\Tmat(n, n^\tau, d) / n^{\tau + \Omega(1)})$, and
    \item worst query time $O(n^{\tau - \Omega(1)} )$.
\end{itemize}

\end{lemma}
\iffalse
\begin{proof}

Let us take an instance from $\mathsf{AHMV}(n,m)$ problem with $M\in \{0,1 \}^{n \times m}$ and $V \in \{0,1\}^{m \times n}$.

We can construct matrix $\mathsf{M} \in \{0,1\}^{n \times 2m}$ and $\mathsf{V} \in \{0,1\}^{2m \times n}$ as follows
\begin{align*}
\mathsf{M} := \begin{bmatrix}
M &
\ov{M}
\end{bmatrix}
\mathrm{~~~and~~~}
\mathsf{V}:= \begin{bmatrix}
V \\
0_{m \times n}
\end{bmatrix}
\end{align*}
where $\ov{M}$ is a matrix that $\ov{M}_{i,j} = 1-M_{i,j}$.

Note that $\| \mathsf{M}_{i,*} \|_1 = m$, for each $i \in [n]$.

Based on the above construction, we will create a new instance $\mathsf{ODAMV}(\wt{n} = n,\wt{m} = 2m,\wt{d}=\wt{m} = 2m)$.

We construct $\wt{Q} \in \R^{ \wt{n} \times \wt{m} }, \wt{K} \in \R^{ \wt{m} \times \wt{m}}, \wt{V} \in \R^{ \wt{m} \times \wt{n}}$ as follows
\begin{align*}
\wt{Q} = \mathsf{M} , ~~~ \wt{K} = I_{\wt{m}}, ~~~ \wt{V} = \mathsf{V}
\end{align*}
From the above construction, we know that
\begin{align*}
\wt{D}_{i,i} = m \exp(1) + m \exp(0) = m (e+1)
\end{align*}

Using Claim~\ref{cla:1_case_AHDAMV}, \ref{cla:empty_between} and Claim~\ref{cla:0_case_AHDAMV}, we know that if we have an algorithm to solve problem $\mathsf{ODAMV}$, then we can solve $\mathsf{AHMV}$.
\end{proof}
\fi
\begin{proof}
Assume there was a dynamic algorithm faster than what is stated in \Cref{lem:lowerbound_D} for some parameter $\tau$, i.e.~update time $O(\Tmat(n, n^\tau, d) / n^{\tau + \epsilon})$ and query time $O(n^{\tau - \epsilon} )$ for some constant $\epsilon > 0$. We show that this would contradict the hinted $Mv$ conjecture (\Cref{con:hintedMv}).

Let us take an instance for the $v$-hinted Mv problem (\Cref{def:hintedMv}) 
with $M\in \{0,1\}^{n \times n}, V\in \{0,1\}^{n \times n}.$

We can construct matrix $\mathsf{M} \in \{0,1\}^{n \times 2n}$ and $\mathsf{V} \in \{0,1\}^{2n \times n}$ as follows
\begin{align*}
\mathsf{M} := \begin{bmatrix}
M &
\ov{M}
\end{bmatrix}
\mathrm{~~~and~~~}
\mathsf{V}:= \begin{bmatrix}
V \\
{\bf 0}_{n \times n}
\end{bmatrix}
\end{align*}
where $\ov{M}$ is a matrix that $\ov{M}_{i,j} = 1-M_{i,j}$.

Note that $\| \mathsf{M}_{i,*} \|_1 = n$, for each $i \in [n]$.

Based on the above construction, we will create a new instance $\mathsf{ODAMV}(\wt{n} = 2n, \wt{d} = 2n)$, where
\begin{align*}
\wt{Q} = 
\begin{bmatrix}
    \mathsf{M}\\
    {\bf 0}_{n \times 2n}
\end{bmatrix} , ~~~ \wt{K} = {\bf 0}_{2n \times 2n}, ~~~ \wt{V} = \begin{bmatrix} \mathsf{V} & {\bf 0}_{2n \times n}
\end{bmatrix}
\end{align*}

During phase \ref{phase1}, we give this input to the dynamic algorithm for the $\mathsf{ODAMV}$ problem (Definition~\ref{def:AHDAMV}).

Let $D \in \{0,1\}^{n \times n}$ denote a diagonal matrix, where $\nnz(D) =   n^\tau $

During phase \ref{phase2}, we receive the $2 n \times 2 n$ diagonal matrix $\mathsf{P} $, where
\begin{align*}
    \mathsf{P} = 
    \begin{bmatrix}
        P & 0 \\
        0 & P 
    \end{bmatrix}
\end{align*}
and  $ \nnz(\mathsf{P}) =2 n^\tau$.

We perform $2n^\tau$ updates to the data structure to set $\wt{K}^\top = \mathsf{P}$. This takes 
\begin{align*}
O(\wt{n}^\tau \cdot (\Tmat(\wt{n}, \wt{n}^\tau, \wt{d}) / \wt{n}^{\tau + \epsilon})) = O(n^{\omega(1,1,\tau)-\epsilon})
\end{align*}
time.

Note that 
\begin{itemize}
\item 
$\| \wt{Q}_{i,*} \|_1 = n$, for each $i \in [n]$.
\item $\| \wt{Q}_{i,*} \|_1 = 0$, for each $i \in [n+1,2n]$.
\end{itemize}
By using the definition of $\mathsf{P}$, we know that, for each $i \in [n]$
\begin{align*}
\wt{D}_{i,i} = {n^\tau} \exp(1) + {n^\tau} \exp(0) =  {n^\tau}(e+1).
\end{align*}
For each $i \in [n+1,2n]$
\begin{align}
\wt{D}_{i,i} = n^{\tau} \exp(0) = n^{\tau}. \label{eq:normalization}
\end{align}
Hence, we don't need to update $\wt{D}$.

At last, in phase \ref{phase3}, we perform $\wt{n}$ queries to obtain the column $\exp(\wt{Q}\wt{K}^\top)\wt{V}_{*,i}$ in $O(\wt{n} \cdot \wt{n}^{\tau-\epsilon}) = O(n^{1+\tau-\epsilon})$ time.

Using Claim~\ref{cla:0_case_AHDAMV} and 
 Claim~\ref{cla:1_case_AHDAMV},
 we know that, for any $i \in [n]$ and for any $j \in [n]$, if there is an algorithm that can find  $(\wt{D}^{-1} \exp( \wt{Q}\wt{K}^\top) \wt{V} )_{j,i}$ , then using $(\wt{D}^{-1} \exp( \wt{Q}\wt{K}^\top) \wt{V} )_{j,i} - (\wt{D}^{-1}\wt{V} )_{j,i}$ is enough to reconstruct $(\mathsf{M}\mathsf{P}\mathsf{V})_{j,i}$. Here $\wt{D}^{-1}\wt{V}$ can be computed in just $O(1)$ time via Eq.~\eqref{eq:normalization}. 
 Thus, we can know the $(MDV)_{j,i}$ for the hinted $Mv$ problem in $O(n^{1+\tau\epsilon})$ time, contradicting the hinted $Mv$ conjecture.

\end{proof}

\begin{claim}\label{cla:1_case_AHDAMV}
For each $i \in [n]$ and $j \in [n]$, if  $(\wt{D}^{-1} ( \exp(\wt{Q} \wt{K}^\top) - {\bf 1}_{\wt{n} \times \wt{n}}) \wt{V})_{j,i} $ is  $> 0$, then $ (\mathsf{M}\mathsf{P} \mathsf{V})_{j,i}  = 1$,
\end{claim}
\begin{proof}
 
By using the fact that $n^\tau(e+1) > 0$ and $n^\tau > 0$, we have
 \begin{align*}
    \wt{D}^{-1} ( \exp(\wt{Q} \wt{K}^\top) - {\bf 1}_{ \wt{n} \times \wt{n} } ) \wt{V})_{j,i} > & ~ 0 \\
    ( (\exp(\wt{Q} \wt{K}^\top) - {\bf 1}_{ \wt{n} \times \wt{n} } ) \wt{V})_{j,i}    > & ~ 0 
 \end{align*}
  We know 
  \begin{align*}
\wt{Q} = 
\begin{bmatrix}
    \mathsf{M}\\
    {\bf 0}_{n \times 2n}
\end{bmatrix} , ~~~ \wt{K}^\top =\begin{bmatrix}
        P & 0 \\
        0 & P
    \end{bmatrix}, ~~~ \wt{V} = \begin{bmatrix} \mathsf{V} & {\bf 0}_{2n \times n}
\end{bmatrix},
\end{align*} 
so we have
 \begin{align*}
      ( (\exp(\mathsf{M} \mathsf{P}) - {\bf 1}_{ n \times 2 n } ) \mathsf{V})_{j,i}  >  & ~  0.
 \end{align*}

 For $k \in [n+1,2n]$, as $\mathsf{V} = \begin{bmatrix}
V \\
{\bf 0}_{n \times n}
\end{bmatrix}$, we know $ (\exp(\mathsf{M}\mathsf{P}) - {\bf 1}_{n \times 2n} )_{j,k} (\mathsf{V})_{k,i} = 0 $.

Using the definition of matrix multiplication, and the fact that $\exp(x)>1$ for all $x>0$, we have some $k \in [n]$ with
 \begin{align*}
       (\exp(\mathsf{M}\mathsf{P}) - {\bf 1}_{n \times 2n} )_{j,k} (\mathsf{V})_{k,i} >  & ~ 0 \\
       (\exp(\mathsf{M}\mathsf{P})_{j,k} - 1) (\mathsf{V})_{k,i}  >  & ~ 0
 \end{align*}
We can conclude that for each $i \in [n], j \in [n]$, there is at least one $k \in [n]$ such that 
\begin{itemize}
    \item $\mathsf{V}_{k,i} > 0$
    \item $(\mathsf{M}\mathsf{P})_{j,k} > 0$
\end{itemize}
Therefore, by using the definition of boolean semi-ring, we can conclude that $(\mathsf{M} \mathsf{P} \mathsf{V})_{j,i}  = 1$

\end{proof}

\begin{claim}\label{cla:0_case_AHDAMV}
For each $i \in [n]$ and $j \in [n]$, if $(\wt{D}^{-1} (\exp(\wt{Q} \wt{K}^\top) - \mathbf{1}_{\wt{n} \times \wt{n}}) \wt{V})_{j,i}$ is $0$ then $(\mathsf{M} \mathsf{P} \mathsf{V})_{j,i}  = 0$.
\end{claim}
\begin{proof}

By using the fact that $n^\tau(e+1) > 0$ and $n^\tau > 0$, we have
 \begin{align*}
    \wt{D}^{-1} ( \exp(\wt{Q} \wt{K}^\top) - {\bf 1}_{ \wt{n} \times \wt{n} } ) \wt{V})_{j,i} = & ~ 0 \\
    ( (\exp(\wt{Q} \wt{K}^\top) - {\bf 1}_{ \wt{n} \times \wt{n} } ) \wt{V})_{j,i}    = & ~ 0 
 \end{align*}

 We know 
  \begin{align*}
\wt{Q} = 
\begin{bmatrix}
    \mathsf{M}\\
    {\bf 0}_{n \times 2n}
\end{bmatrix} , ~~~ \wt{K}^\top =\begin{bmatrix}
        P & 0 \\
        0 & P 
    \end{bmatrix}, ~~~ \wt{V} = \begin{bmatrix} \mathsf{V} & {\bf 0}_{2n \times n}
\end{bmatrix},
\end{align*} 
so we have
 \begin{align*}
      ( (\exp(\mathsf{M} \mathsf{P}) - {\bf 1}_{ n \times 2 n } ) \mathsf{V})_{j,i}  =  & ~  0.
 \end{align*}

For $k \in [n+1,2n]$, as $\mathsf{V} = \begin{bmatrix}
V \\
{\bf 0}_{n \times n}
\end{bmatrix}$, we know $ (\exp(\mathsf{M}\mathsf{P}) - {\bf 1}_{n \times 2n} )_{j,k} (\mathsf{V})_{k,i} = 0 $.

For all $k \in [n]$ such that $\mathsf{V}_{k,i }  =  1$ , we have $(\exp(\mathsf{M}\mathsf{P}) - \mathbf{1}_{n \times 2n})_{j,k} = 0$ , which also implies that $(\mathsf{M} \mathsf{P})_{j,k} = 0$.

 Now, we can conclude that  $(\mathsf{M}\mathsf{P} \mathsf{V})_{j,i}  = 0$ for each $i \in [n]$ and $j \in [n]$.
\end{proof}

\ifdefined\isarxiv
%\section*{Acknowledgments}
\bibliographystyle{alpha}
\bibliography{ref}
\else
\bibliography{ref}

\newcommand{\etalchar}[1]{$^{#1}$}
\begin{thebibliography}{WLK{\etalchar{+}}20}

\bibitem[ACSS20]{acss20}
Josh Alman, Timothy Chu, Aaron Schild, and Zhao Song.
\newblock Algorithms and hardness for linear algebra on geometric graphs.
\newblock In {\em 2020 IEEE 61st Annual Symposium on Foundations of Computer
  Science (FOCS)}, pages 541--552. IEEE, 2020.

\bibitem[AS23]{as23}
Josh Alman and Zhao Song.
\newblock Fast attention requires bounded entries.
\newblock {\em arXiv preprint arXiv:2302.13214}, 2023.

\bibitem[AW21]{aw21}
Josh Alman and Virginia~Vassilevska Williams.
\newblock A refined laser method and faster matrix multiplication.
\newblock In {\em Proceedings of the 2021 ACM-SIAM Symposium on Discrete
  Algorithms (SODA)}, pages 522--539. SIAM, 2021.

\bibitem[BAG20]{bag20}
Satwik Bhattamishra, Kabir Ahuja, and Navin Goyal.
\newblock On the {A}bility and {L}imitations of {T}ransformers to {R}ecognize
  {F}ormal {L}anguages.
\newblock In {\em Proceedings of the 2020 Conference on Empirical Methods in
  Natural Language Processing (EMNLP)}, pages 7096--7116, Online, November
  2020. Association for Computational Linguistics.

\bibitem[Bel22]{b22}
Yonatan Belinkov.
\newblock Probing classifiers: Promises, shortcomings, and advances.
\newblock {\em Computational Linguistics}, 48(1):207--219, March 2022.

\bibitem[BIS17]{bis17}
Arturs Backurs, Piotr Indyk, and Ludwig Schmidt.
\newblock On the fine-grained complexity of empirical risk minimization: Kernel
  methods and neural networks.
\newblock {\em Advances in Neural Information Processing Systems}, 30, 2017.

\bibitem[BLSS20]{blss20}
Jan van~den Brand, Yin~Tat Lee, Aaron Sidford, and Zhao Song.
\newblock Solving tall dense linear programs in nearly linear time.
\newblock In {\em Proceedings of the 52nd Annual ACM SIGACT Symposium on Theory
  of Computing}, pages 775--788, 2020.

\bibitem[BMR{\etalchar{+}}20]{bmr+20}
Tom Brown, Benjamin Mann, Nick Ryder, Melanie Subbiah, Jared~D Kaplan, Prafulla
  Dhariwal, Arvind Neelakantan, Pranav Shyam, Girish Sastry, Amanda Askell,
  et~al.
\newblock Language models are few-shot learners.
\newblock {\em Advances in neural information processing systems},
  33:1877--1901, 2020.

\bibitem[BN19]{bn19}
Jan van~den Brand and Danupon Nanongkai.
\newblock Dynamic approximate shortest paths and beyond: Subquadratic and
  worst-case update time.
\newblock In {\em 2019 IEEE 60th Annual Symposium on Foundations of Computer
  Science (FOCS)}, pages 436--455. IEEE, 2019.

\bibitem[BNS19]{bns19}
Jan van~den Brand, Danupon Nanongkai, and Thatchaphol Saranurak.
\newblock Dynamic matrix inverse: Improved algorithms and matching conditional
  lower bounds.
\newblock In {\em 2019 IEEE 60th Annual Symposium on Foundations of Computer
  Science (FOCS)}, pages 456--480. IEEE, 2019.

\bibitem[BPG20]{bpg20}
Satwik Bhattamishra, Arkil Patel, and Navin Goyal.
\newblock On the computational power of transformers and its implications in
  sequence modeling.
\newblock In {\em Proceedings of the 24th Conference on Computational Natural
  Language Learning}, pages 455--475, Online, November 2020. Association for
  Computational Linguistics.

\bibitem[Bra20]{b20}
Jan van~den Brand.
\newblock A deterministic linear program solver in current matrix
  multiplication time.
\newblock In {\em Proceedings of the Fourteenth Annual ACM-SIAM Symposium on
  Discrete Algorithms (SODA)}, pages 259--278. SIAM, 2020.

\bibitem[Bra21]{b21}
Jan van~den Brand.
\newblock Unifying matrix data structures: Simplifying and speeding up
  iterative algorithms.
\newblock In {\em Symposium on Simplicity in Algorithms (SOSA)}, pages 1--13.
  SIAM, 2021.

\bibitem[CKL18]{ckl18}
Diptarka Chakraborty, Lior Kamma, and Kasper~Green Larsen.
\newblock Tight cell probe bounds for succinct boolean matrix-vector
  multiplication.
\newblock In {\em Proceedings of the 50th Annual ACM SIGACT Symposium on Theory
  of Computing (STOC)}, pages 1297--1306, 2018.

\bibitem[CKLM19]{cklm19}
Kevin Clark, Urvashi Khandelwal, Omer Levy, and Christopher~D. Manning.
\newblock What does {BERT} look at? an analysis of {BERT}{'}s attention.
\newblock In {\em Proceedings of the 2019 ACL Workshop BlackboxNLP: Analyzing
  and Interpreting Neural Networks for NLP}, pages 276--286, Florence, Italy,
  August 2019. Association for Computational Linguistics.

\bibitem[CKNS20]{ckns20}
Moses Charikar, Michael Kapralov, Navid Nouri, and Paris Siminelakis.
\newblock Kernel density estimation through density constrained near neighbor
  search.
\newblock In {\em 2020 IEEE 61st Annual Symposium on Foundations of Computer
  Science (FOCS)}, pages 172--183. IEEE, 2020.

\bibitem[CLP{\etalchar{+}}21]{clp+21}
Beidi Chen, Zichang Liu, Binghui Peng, Zhaozhuo Xu, Jonathan~Lingjie Li, Tri
  Dao, Zhao Song, Anshumali Shrivastava, and Christopher Re.
\newblock Mongoose: A learnable lsh framework for efficient neural network
  training.
\newblock In {\em International Conference on Learning Representations}, 2021.

\bibitem[CLS19]{cls19}
Michael~B Cohen, Yin~Tat Lee, and Zhao Song.
\newblock Solving linear programs in the current matrix multiplication time.
\newblock In {\em STOC}, 2019.

\bibitem[CND{\etalchar{+}}22]{cnd+22}
Aakanksha Chowdhery, Sharan Narang, Jacob Devlin, Maarten Bosma, Gaurav Mishra,
  Adam Roberts, Paul Barham, Hyung~Won Chung, Charles Sutton, Sebastian
  Gehrmann, et~al.
\newblock Palm: Scaling language modeling with pathways.
\newblock {\em arXiv preprint arXiv:2204.02311}, 2022.

\bibitem[DCLT18]{dclt18}
Jacob Devlin, Ming-Wei Chang, Kenton Lee, and Kristina Toutanova.
\newblock Bert: Pre-training of deep bidirectional transformers for language
  understanding.
\newblock {\em arXiv preprint arXiv:1810.04805}, 2018.

\bibitem[DI00]{di00}
Camil Demetrescu and Giuseppe~F Italiano.
\newblock Fully dynamic transitive closure: breaking through the o (n/sup 2/)
  barrier.
\newblock In {\em Proceedings 41st Annual Symposium on Foundations of Computer
  Science}, pages 381--389. IEEE, 2000.

\bibitem[DLY21]{dly21}
Sally Dong, Yin~Tat Lee, and Guanghao Ye.
\newblock A nearly-linear time algorithm for linear programs with small
  treewidth: A multiscale representation of robust central path.
\newblock In {\em Proceedings of the 53rd Annual ACM SIGACT Symposium on Theory
  of Computing}, pages 1784--1797, 2021.

\bibitem[EGKZ22]{egkz22}
Benjamin~L Edelman, Surbhi Goel, Sham Kakade, and Cyril Zhang.
\newblock Inductive biases and variable creation in self-attention mechanisms.
\newblock In Kamalika Chaudhuri, Stefanie Jegelka, Le~Song, Csaba Szepesvari,
  Gang Niu, and Sivan Sabato, editors, {\em Proceedings of the 39th
  International Conference on Machine Learning}, volume 162 of {\em Proceedings
  of Machine Learning Research}, pages 5793--5831. PMLR, 17--23 Jul 2022.

\bibitem[EGZ20]{egz20}
Javid Ebrahimi, Dhruv Gelda, and Wei Zhang.
\newblock How can self-attention networks recognize {D}yck-n languages?
\newblock In {\em Findings of the Association for Computational Linguistics:
  EMNLP 2020}, pages 4301--4306, Online, November 2020. Association for
  Computational Linguistics.

\bibitem[GLSS18]{glss18}
Ankit Garg, Yin~Tat Lee, Zhao Song, and Nikhil Srivastava.
\newblock A matrix expander chernoff bound.
\newblock In {\em Proceedings of the 50th Annual ACM SIGACT Symposium on Theory
  of Computing}, pages 1102--1114, 2018.

\bibitem[GR21]{gr21}
Yong Gu and Hanlin Ren.
\newblock Constructing a distance sensitivity oracle in $ o (n^{2.5794} m)$
  time.
\newblock {\em arXiv preprint arXiv:2102.08569}, 2021.

\bibitem[GS22]{gs22}
Yuzhou Gu and Zhao Song.
\newblock A faster small treewidth sdp solver.
\newblock {\em arXiv preprint arXiv:2211.06033}, 2022.

\bibitem[HJS{\etalchar{+}}22]{hjs+22}
Baihe Huang, Shunhua Jiang, Zhao Song, Runzhou Tao, and Ruizhe Zhang.
\newblock Solving sdp faster: A robust ipm framework and efficient
  implementation.
\newblock In {\em 2022 IEEE 63rd Annual Symposium on Foundations of Computer
  Science (FOCS)}, pages 233--244. IEEE, 2022.

\bibitem[HKNS15]{hkns15}
Monika Henzinger, Sebastian Krinninger, Danupon Nanongkai, and Thatchaphol
  Saranurak.
\newblock Unifying and strengthening hardness for dynamic problems via the
  online matrix-vector multiplication conjecture.
\newblock In {\em Proceedings of the forty-seventh annual ACM symposium on
  Theory of computing (STOC)}, pages 21--30, 2015.

\bibitem[HL19]{hl19}
John Hewitt and Percy Liang.
\newblock Designing and interpreting probes with control tasks.
\newblock In {\em Proceedings of the 2019 Conference on Empirical Methods in
  Natural Language Processing and the 9th International Joint Conference on
  Natural Language Processing (EMNLP-IJCNLP)}, pages 2733--2743, Hong Kong,
  China, November 2019. Association for Computational Linguistics.

\bibitem[HM19]{hm19}
John Hewitt and Christopher~D. Manning.
\newblock {A} structural probe for finding syntax in word representations.
\newblock In {\em Proceedings of the 2019 Conference of the North {A}merican
  Chapter of the Association for Computational Linguistics: Human Language
  Technologies, Volume 1 (Long and Short Papers)}, pages 4129--4138,
  Minneapolis, Minnesota, June 2019. Association for Computational Linguistics.

\bibitem[JKL{\etalchar{+}}20]{jkl+20}
Haotian Jiang, Tarun Kathuria, Yin~Tat Lee, Swati Padmanabhan, and Zhao Song.
\newblock A faster interior point method for semidefinite programming.
\newblock In {\em 2020 IEEE 61st annual symposium on foundations of computer
  science (FOCS)}, pages 910--918. IEEE, 2020.

\bibitem[JLSW20]{jlsw20}
Haotian Jiang, Yin~Tat Lee, Zhao Song, and Sam Chiu-wai Wong.
\newblock An improved cutting plane method for convex optimization,
  convex-concave games, and its applications.
\newblock In {\em Proceedings of the 52nd Annual ACM SIGACT Symposium on Theory
  of Computing}, pages 944--953, 2020.

\bibitem[JSWZ21]{jswz21}
Shunhua Jiang, Zhao Song, Omri Weinstein, and Hengjie Zhang.
\newblock Faster dynamic matrix inverse for faster lps.
\newblock In {\em STOC}. arXiv preprint arXiv:2004.07470, 2021.

\bibitem[KKL20]{kkll20}
Nikita Kitaev, {\L}ukasz Kaiser, and Anselm Levskaya.
\newblock Reformer: The efficient transformer.
\newblock {\em arXiv preprint arXiv:2001.04451}, 2020.

\bibitem[KVPF20]{kvpf20}
Angelos Katharopoulos, Apoorv Vyas, Nikolaos Pappas, and Fran{\c{c}}ois
  Fleuret.
\newblock Transformers are rnns: Fast autoregressive transformers with linear
  attention.
\newblock In {\em International Conference on Machine Learning}, pages
  5156--5165. PMLR, 2020.

\bibitem[LG14]{lg14}
Fran\c{c}ois Le~Gall.
\newblock Powers of tensors and fast matrix multiplication.
\newblock In {\em Proceedings of the 39th International Symposium on Symbolic
  and Algebraic Computation}, ISSAC '14, 2014.

\bibitem[LLR23]{llr23}
Yuchen Li, Yuanzhi Li, and Andrej Risteski.
\newblock How do transformers learn topic structure: Towards a mechanistic
  understanding.
\newblock {\em arXiv preprint arXiv:2303.04245}, 2023.

\bibitem[LSZ19]{lsz19}
Yin~Tat Lee, Zhao Song, and Qiuyi Zhang.
\newblock Solving empirical risk minimization in the current matrix
  multiplication time.
\newblock In {\em COLT}, 2019.

\bibitem[LSZ23]{lsz23}
Zhihang Li, Zhao Song, and Tianyi Zhou.
\newblock Solving regularized exp, cosh and sinh regression problem.
\newblock {\em arxiv preprint 2303.15725}, 2023.

\bibitem[LW17]{lw17}
Kasper~Green Larsen and Ryan Williams.
\newblock Faster online matrix-vector multiplication.
\newblock In {\em Proceedings of the Twenty-Eighth Annual ACM-SIAM Symposium on
  Discrete Algorithms (SODA)}, pages 2182--2189, 2017.

\bibitem[Ope23]{openai23}
OpenAI.
\newblock Gpt-4 technical report, 2023.

\bibitem[PMB19]{pmb19}
Jorge P{\'e}rez, Javier Marinkovi{\'c}, and Pablo Barcel{\'o}.
\newblock On the turing completeness of modern neural network architectures.
\newblock {\em arXiv preprint arXiv:1901.03429}, 2019.

\bibitem[RNS{\etalchar{+}}18]{rns+18}
Alec Radford, Karthik Narasimhan, Tim Salimans, Ilya Sutskever, et~al.
\newblock Improving language understanding by generative pre-training.
\newblock 2018.

\bibitem[San04]{s04}
Piotr Sankowski.
\newblock Dynamic transitive closure via dynamic matrix inverse.
\newblock In {\em 45th Annual IEEE Symposium on Foundations of Computer
  Science}, pages 509--517. IEEE, 2004.

\bibitem[San05]{s05}
Piotr Sankowski.
\newblock Subquadratic algorithm for dynamic shortest distances.
\newblock In {\em Computing and Combinatorics: 11th Annual International
  Conference, COCOON 2005 Kunming, China, August 16--19, 2005 Proceedings 11},
  pages 461--470. Springer, 2005.

\bibitem[SS92]{ss92}
Hava~T. Siegelmann and Eduardo~D. Sontag.
\newblock On the computational power of neural nets.
\newblock In {\em Proceedings of the Fifth Annual Workshop on Computational
  Learning Theory}, COLT '92, page 440–449, New York, NY, USA, 1992.
  Association for Computing Machinery.

\bibitem[SY21]{sy21}
Zhao Song and Zheng Yu.
\newblock Oblivious sketching-based central path method for solving linear
  programming problems.
\newblock In {\em 38th International Conference on Machine Learning (ICML)},
  2021.

\bibitem[TDP19]{tdp19}
Ian Tenney, Dipanjan Das, and Ellie Pavlick.
\newblock {BERT} rediscovers the classical {NLP} pipeline.
\newblock In {\em Proceedings of the 57th Annual Meeting of the Association for
  Computational Linguistics}, pages 4593--4601, Florence, Italy, July 2019.
  Association for Computational Linguistics.

\bibitem[VB19]{vb19}
Jesse Vig and Yonatan Belinkov.
\newblock Analyzing the structure of attention in a transformer language model.
\newblock In {\em Proceedings of the 2019 ACL Workshop BlackboxNLP: Analyzing
  and Interpreting Neural Networks for NLP}, pages 63--76, Florence, Italy,
  August 2019. Association for Computational Linguistics.

\bibitem[VSP{\etalchar{+}}17]{vsp+17}
Ashish Vaswani, Noam Shazeer, Niki Parmar, Jakob Uszkoreit, Llion Jones,
  Aidan~N Gomez, {\L}ukasz Kaiser, and Illia Polosukhin.
\newblock Attention is all you need.
\newblock {\em Advances in neural information processing systems}, 30, 2017.

\bibitem[WCM21]{wcm21}
Colin Wei, Yining Chen, and Tengyu Ma.
\newblock Statistically meaningful approximation: a case study on approximating
  turing machines with transformers, 2021.

\bibitem[Wil12]{w12}
Virginia~Vassilevska Williams.
\newblock Multiplying matrices faster than coppersmith-winograd.
\newblock In {\em Proceedings of the forty-fourth annual ACM symposium on
  Theory of computing (STOC)}, pages 887--898. ACM, 2012.

\bibitem[WLK{\etalchar{+}}20]{wlk+20}
Sinong Wang, Belinda~Z Li, Madian Khabsa, Han Fang, and Hao Ma.
\newblock Linformer: Self-attention with linear complexity.
\newblock {\em arXiv preprint arXiv:2006.04768}, 2020.

\bibitem[YBR{\etalchar{+}}20]{ybr+20}
Chulhee Yun, Srinadh Bhojanapalli, Ankit~Singh Rawat, Sashank Reddi, and Sanjiv
  Kumar.
\newblock Are transformers universal approximators of sequence-to-sequence
  functions?
\newblock In {\em International Conference on Learning Representations}, 2020.

\bibitem[YPPN21]{yppn21}
Shunyu Yao, Binghui Peng, Christos Papadimitriou, and Karthik Narasimhan.
\newblock Self-attention networks can process bounded hierarchical languages.
\newblock In {\em Proceedings of the 59th Annual Meeting of the Association for
  Computational Linguistics and the 11th International Joint Conference on
  Natural Language Processing (Volume 1: Long Papers)}, pages 3770--3785,
  Online, August 2021. Association for Computational Linguistics.

\bibitem[ZBB{\etalchar{+}}22]{zbb+22}
Yi~Zhang, Arturs Backurs, Sébastien Bubeck, Ronen Eldan, Suriya Gunasekar, and
  Tal Wagner.
\newblock Unveiling transformers with lego: a synthetic reasoning task, 2022.

\bibitem[ZBIW19]{zbiw19}
Ofir Zafrir, Guy Boudoukh, Peter Izsak, and Moshe Wasserblat.
\newblock Q8bert: Quantized 8bit bert.
\newblock In {\em 2019 Fifth Workshop on Energy Efficient Machine Learning and
  Cognitive Computing-NeurIPS Edition (EMC2-NIPS)}, pages 36--39. IEEE, 2019.

\bibitem[ZHDK23]{zhdk23}
Amir Zandieh, Insu Han, Majid Daliri, and Amin Karbasi.
\newblock Kdeformer: Accelerating transformers via kernel density estimation.
\newblock {\em arXiv preprint arXiv:2302.02451}, 2023.

\bibitem[ZPGA23]{zpga23}
Haoyu Zhao, Abhishek Panigrahi, Rong Ge, and Sanjeev Arora.
\newblock Do transformers parse while predicting the masked word?
\newblock {\em arXiv preprint arXiv:2303.08117}, 2023.

\bibitem[ZRG{\etalchar{+}}22]{zrg+22}
Susan Zhang, Stephen Roller, Naman Goyal, Mikel Artetxe, Moya Chen, Shuohui
  Chen, Christopher Dewan, Mona Diab, Xian Li, Xi~Victoria Lin, et~al.
\newblock Opt: Open pre-trained transformer language models.
\newblock {\em arXiv preprint arXiv:2205.01068}, 2022.

\bibitem[Zwi02]{z02}
Uri Zwick.
\newblock All pairs shortest paths using bridging sets and rectangular matrix
  multiplication.
\newblock {\em Journal of the ACM (JACM)}, 49(3):289--317, 2002.

\end{thebibliography}
\bibliographystyle{alpha}

\fi

\newpage
\onecolumn
\appendix

%%%% Cut-line between first 10 pages and appendix

%%% some writing rules

%% Writing rule for creating tags.
%% Tags :
%% Theorem    \ref{thm:bla_bla}
%% Lemma      \ref{lem:bla_bla}
%% Claim      \ref{cla:bla_bla}
%% Corollary  \ref{cor:bla_bla}
%% Fact       \ref{fac:bla_bla}
%% Definition \ref{def:bla_bla}
%% Section    \ref{sec:bla_bla}
%% Subsection \ref{sub:bla_bla}
%% Equation   \ref{eq:bla_bla}

\end{document}